\documentclass[nofootinbib,notitlepage,superscriptaddress,twocolumn,colorlinks=true,linkcolor=blue,citecolor=blue,urlcolor=blue]{revtex4-2}
\pdfoutput=1

\usepackage{aas_macros,amsmath,amssymb,esint,graphicx,overpic,xcolor,mathrsfs,stackengine,array}
\usepackage[bottom]{footmisc}

\usepackage{comment}
\usepackage{orcidlink}

\usepackage{tensor}
\usepackage{physics}
\usepackage{multirow}
\usepackage{ulem}
\usepackage{hyperref}

\newcommand{\bs}{\begin{split}}
\newcommand{\es}{\end{split}}

\newcommand{\R}{\mathbb{R}}

\let\originalleft\left
\let\originalright\right
\renewcommand{\left}{\mathopen{}\mathclose\bgroup\originalleft}
\renewcommand{\right}{\aftergroup\egroup\originalright}

\newcolumntype{P}[1]{>{\hspace{5pt}}p{#1}<{\hspace{5pt}}}

\usepackage{booktabs}
\usepackage{hyperref}
\hypersetup{
	colorlinks=true,
	linkcolor=blue,
	citecolor=cyan,
	urlcolor=cyan,
}

\begin{document}

\title{Testing Conformal Weyl Gravity with Plasma Modified Black Hole Shadows and EHT Constraints}

\author{Zhichen Guan}
\affiliation{Department of Mechanical Engineering, Johns Hopkins University, Baltimore, MD 21218}

\author{Connor McMillin}
\affiliation{Department of Physics, Montana State University, Bozeman, MT 59717}

\author{Owen Gartlan}
\affiliation{Department of Physics, Grinnell College, Grinnell, IA 50112}
\author{William Horner}
\affiliation{College of Law, University of Iowa, Iowa City, IA 52242}

\author{Leo Rodriguez}
\affiliation{Department of Physics, Grinnell College, Grinnell, IA 50112}

\author{Shanshan Rodriguez\orcidlink{0000-0003-2944-0424}}
\email{rodriguezs@grinnell.edu}
\thanks{Corresponding author} 
\affiliation{Center for Astrophysics $\arrowvert$ Harvard \& Smithsonian, 60 Garden Street, Cambridge, MA 02138}
\affiliation{Department of Physics, Grinnell College, Grinnell, IA 50112}

\begin{abstract}
We study the shadows of a charged rotating de Sitter solution in fourth order conformal Weyl gravity (CWG) immersed in plasma and analyze the joint charge-plasma constraints. Our main goal is to distinguish observational effects caused by the dispersive plasma environment from those caused by the underlying gravitational theory. Within a separable Hamilton-Jacobi plasma model, we derive the radial and angular photon potentials, the equatorial finite-distance deflection angle, and the finite-observer shadow boundary due to the asymptotically non-flat nature of the CWG spacetime. Our results separate two effects. Increasing the plasma strength generally reduces the shadow radius, the signed deflection angle, and the energy emission peak in both CWG and GR. Charge ($q$) produces a stronger theory dependence: for the representative parameters studied, increasing $q$ enlarges the CWG shadow but reduces the GR shadow. The lensing behavior independently supports the geometry-dependent response found from the photon region and shadow observables. We further compare the finite observer dimensionless shadow radius with EHT-inferred size intervals for Sgr~A$^*$ and M87$^*$. Our results indicate that plasma effects can mimic part of the shadow size shift, but the distinct charge response of CWG offers a potential observational route for separating conformal gravity signatures from environmental propagation effects.\\

\end{abstract}
\maketitle
\section{Introduction}\label{sec:1}
Black hole shadows and strong field gravitational lensing have become powerful observational probes of gravity and the near horizon environment of compact objects. The theoretical basis for interpreting these images goes back to the photon region and its projection on the observer's sky, while modern shadow observables quantify both the characteristic radius and deviations from circularity \cite{Synge1966,Bardeen1973,Hioki:2009na,Abdujabbarov:2015xqa,PerlickTsupko2022Review}. The first horizon scale image of the supermassive black hole in M87* released by the Event Horizon Telescope (EHT) revealed a bright asymmetric ring surrounding a central brightness depression whose angular scale is consistent with the characteristic photon orbit size predicted by general relativity. The subsequent image of Sgr A* extended these tests to the Galactic center, where rapid source variability and interstellar scattering introduce additional observational complications \cite{EventHorizonTelescope:2019dse,EventHorizonTelescope2019VPhysicalOrigin,EventHorizonTelescope:2022wkp,EventHorizonTelescope:2022wok}. These observations motivate theoretical studies of how both spacetime geometry and astrophysical environments modify the apparent size, deformation, and lensing structure of black hole shadows. Recent imaging forecasts further emphasize that current and next-generation black hole shadow measurements can test gravity only when geometric deviations are analyzed together with image-formation systematics \cite{UniyalDihingiaMizunoRezzolla2026}. 

In realistic black hole environments, light does not propagate through vacuum alone. Plasma surrounding the compact object introduces refraction and dispersion, so the shadow boundary, deflection angle, and associated observables become frequency dependent. The standard analytical treatment of light propagation in a non-magnetized, pressureless plasma replaces the vacuum null condition by the Hamiltonian $H=(g^{\mu\nu}p_\mu p_\nu+\omega_p^2)/2$, where $\omega_p$ is the local plasma frequency. Equivalently, the photon propagation can be described through an effective refractive index satisfying $n^2=1-\omega_p^2/\omega^2$, where $\omega$ is the photon frequency measured by the local observer. The condition $\omega>\omega_p$ determines the range in which the refractive index is real and light can propagate through the medium. Inhomogeneous plasma produces an additional refractive contribution to the gravitational bending angle, while homogeneous plasma still modifies photon propagation through the gravitational redshift of the local photon frequency \cite{Morozova:2013uyv,BisnovatyiKoganTsupko2010,TsupkoBisnovatyiKogan2013,BisnovatyiKoganTsupko2017,Ozel:2021ayr}. This framework provides the basis for comparing vacuum shadows with plasma-modified shadows in both general relativity and modified gravity.

Analytical studies of plasma-modified shadows began with models in which the medium preserves enough symmetry for the photon equations to remain tractable. For static, spherically symmetric black holes, the angular shadow radius can be written for a general radial plasma density profile; this result established that the shadow is not only frequency dependent, but also sensitive to whether the plasma density is homogeneous or spatially varying \cite{PerlickTsupkoBisnovatyiKogan2015,Huang:2018rfn}. Power law plasma profiles around Schwarzschild and Kerr black holes showed that the photon sphere need not move in the same way as the apparent shadow, so that the observed shadow size is a combined effect of gravitational redshift, refraction, spin, inclination, and the electron density profile \cite{AtamurotovAhmedovAbdujabbarov2015}. Regular rotating black holes embedded in plasma demonstrated that deformation parameters and plasma terms can both change the radius and distortion of the shadow, producing a degeneracy between geometry and medium even before one introduces a specific modified gravity theory \cite{AbdujabbarovAmirAhmedovGhosh2016}. In Kerr spacetime the separability of the Hamilton-Jacobi equation survives only when the plasma frequency has the form $\omega_p^2=[f_r(r)+f_\theta(\theta)]/\rho^2$, which gives analytic photon regions, generalized Carter constants, and celestial coordinates for observers in a dispersive medium \cite{Perlick2017PlasmaKerr}. Kerr-Newman shadows in plasma were then constructed with the same Hamilton-Jacobi method, showing how charge, spin, plasma strength, and inclination jointly enter the effective potential, the shadow boundary, and the weak deflection angle \cite{BabarBabarAtamurotov2020}. A high spin Kerr black hole surrounded by plasma displays frequency-dependent changes in the shadow and in orbiting hot spot observables, emphasizing that plasma affects not only the boundary of the dark region but also the broader interferometric image structure \cite{Yan2019HighSpinKerrPlasma}. The same separability logic was extended to classes of axisymmetric, stationary black holes generated by modified Newman-Janis algorithms, allowing Kerr-Newman-like, braneworld, Horndeski, and Einstein-Gauss-Bonnet-inspired solutions to be treated within one analytic plasma-shadow framework \cite{BadiaEiroa2021}. 

Asymptotically flat Kerr-like black holes in GR, effective matter models, and modified-gravity extensions have been studied most extensively. This class includes Kerr, Kerr-Newman, Kerr-Sen-like, regular, and dark-matter-embedded rotating black holes, where homogeneous, power-law, Gaussian, Shapiro-type, magnetized, and axion-plasmon profiles have been used to quantify how plasma changes the shadow radius, deformation, photon-region structure, and weak deflection angle  \cite{AtamurotovAhmedovAbdujabbarov2015,DastanSaffariSoroushfar2016KerrSenPlasma,
AbdujabbarovAmirAhmedovGhosh2016,Perlick2017PlasmaKerr,Yan2019HighSpinKerrPlasma,
BabarBabarAtamurotov2020,FathiOlivaresVillanueva2021Plasma,
SarikulovAtamurotovAbdujabbarovAhmedov2022KerrLikePlasma,
ZhangYanGuoChen2022GaussianPlasma,DasSahaGangopadhyay2022PFDMPlasma,
Khodadi2022AxionPlasmon,KumarUniyalChakrabarti2024PlasmaSpacetime,
FengYangChen2024KerrSenPlasma,Raza:2024zkp,YasminJafarzadeJamil2025,
Ali:2025dje,Kala2025HorndeskiPlasma,SahooBanerjee2025KerrSenPlasma,McMillin:2026tok}.
Recent EHT-oriented plasma studies have extended this program to asymptotically safe, Horndeski, Einstein-Maxwell-dilaton, braneworld, hairy, and regular $R^2$ Gaussian black holes, strengthening the point that plasma and non-Kerr geometry must be constrained together rather than separately \cite{Turakhonov2025ASGPlasma,Hoshimov2025HorndeskiPlasma,Ghaderi2026EMDPlasmaEHT,SahooBanerjee2025BraneworldPlasma,Bakhodirov2025HairyPlasma,PradhanGhaderiZeyauddin2026R2Gaussian}. 

In addition, plasma influences on the shadow, deflection angle and gravitational lensing also depends on the asymptotic structure. In asymptotically flat spacetimes, the observer is usually placed at spatial infinity, the celestial coordinates are normalized there, and the impact parameters can be read directly from the constants of motion. For dS and AdS black holes, this simplification is generally not available: de Sitter geometries contain a cosmological horizon, AdS geometries do not approach Minkowski space, and the shadow must be projected onto a local observer screen at finite radius. Analytical shadow studies therefore use local tetrads, finite distance angular radii, or observer dependent celestial coordinates, and the plasma frequency must be evaluated with the locally measured photon frequency rather than with an unambiguous frequency at infinity
\cite{PerlickTsupko2022Review,ChowdhuriBhattacharyya2021PlasmaShadow}. Charged or accelerating AdS examples show that the plasma contribution must be combined with the cosmological curvature rather than added as an asymptotically flat correction \cite{Das:2022tqi,MandalUpadhyayMyrzakulovYergaliyeva2022RNAdSPlasma, AliTiechengAwaisBabar2024AdSPlasma,PradhanGhaderiZeyauddin2026RNAdSPlasma}. Higher curvature and scale dependent examples show that the sign and magnitude of the cosmological term can change how plasma and deformation parameters affect the shadow \cite{Das:2019sty, Ovgun:2023ego}. The weak deflection angle is similarly more delicate in non-asymptotically flat spacetimes, because the usual source lens observer configuration at infinity has to be replaced by a finite distance prescription, often formulated through the optical metric and the Gauss-Bonnet theorem \cite{IshiharaSuzukiOnoKitamuraAsada2016,OnoAsada2019FiniteDistance,
OnoAsada2020NonFlatDeflection,Crisnejo:2018uyn,Crisnejo:2018ppm,Feleppa:2024vdk}. 

The recurring conclusion across these groups is that plasma effects are not a minor correction to the vacuum shadow; they can imitate or mask changes caused by spin, charge, scalar hair, dark matter, Lorentz violation, higher curvature terms, or phenomenological deformation parameters. A shadow constraint on a non-Kerr metric is therefore incomplete unless the plasma profile is included as part of the spacetime optics model. Conformal Weyl gravity (CWG) provides a natural setting in which to study the separation of such plasma-modified optical effects from the GR spacetimes. The theory is constructed from the Weyl-squared action, leading to fourth order field equations that are invariant under local conformal transformations. In the static and spherically symmetric sector, the Mannheim-Kazanas solution contains the Schwarzschild potential, a linear potential term, and a cosmological term. The linear term has been used in phenomenological discussions of galactic rotation curves, while the black hole provides a framework for studying how higher derivative gravity changes strong field observables \cite{Mannheim:1988dj}. 

Charged and rotating CWG black holes differ from their general relativistic counterparts through the radial metric function, where the charge contribution enters in a form that is not equivalent to the Kerr Newman de Sitter expression \cite{MannheimKazanas1991,LiuLu2013}. Since the radial function controls the horizon structure and photon region, the CWG charge term is expected to leave imprints on lensing and shadow observables. Vacuum shadow studies in fourth-order conformal Weyl gravity have shown that the additional CWG parameters can change the apparent shadow size of rotating neutral black holes, while rotating charged Weyl black holes exhibit modified ergosphere, photon region, and shadow structures \cite{MureikaVarieschi2017,AliAmir2019,Meng2022PhotonRegions,QuWangFeng2024ConformalImages,HeydariFard2025CWGAppearance}. Related Weyl-geometric and Weyl-connection models have also been tested with shadow and strong-lensing observables in the last two years \cite{Molla2025WeylGeometric,Gomes2026WeylConnection}. In addition, for scalar perturbations, the modified charge dependence can also suppress superradiant amplification relative to the Kerr Newman de Sitter case over the parameter ranges studied \cite{Gartlan:2026wdz} due to the impact of the charge term of the radial function on the ergoregion and effective potentials.

The current CWG shadow literature either considers a neutral Weyl black hole, a differently parameterized rotating charged Weyl geometry, or vacuum photon motion. In this work, we extend the study to include the influence of non-magnetized, pressureless plasma on the shadow and gravitational lensing of a rotating charged de Sitter black hole in conformal Weyl gravity. We consider a more general inhomogeneous plasma profile, derive the photon equations of motion from the plasma Hamiltonian, and analyze the corresponding effective potentials, deflection angle, shadow radius, and deformation. We compare the plasma-modified CWG results with the vacuum limit and with the general relativistic case in order to identify which features arise from the CWG charge, spin, and cosmological parameters and which are produced by the plasma distribution.

The paper is organized as follows. In Section II, we review the rotating charged de Sitter CWG black hole and its horizon structure. In Section III, we introduce the plasma model and Hamiltonian framework for photon motion. In Sections IV-VI, we study plasma effects on the deflection angle, black hole shadow and its observables, and energy emissions. In Section VII, we compare the combined effects by plasma and charge between two theories and discuss their relevance for EHT-scale observations. We summarize the results in Section VIII. We use geometrized units $c=G=1$ throughout.
\section{Rotating charged (A)dS black hole in conformal Weyl gravity}\label{sec:2}
The charged rotating de Sitter black hole we consider is a solution in Conformal Weyl Gravity (CWG), with an action minimally coupled to the Maxwell Field given by
\begin{gather}
S=\alpha \int d^4x\sqrt{-g}\left( \frac{1}{2} C^{\mu \nu \rho \sigma} C_{\mu \nu \rho \sigma} + \frac{1}{3} F^2 \right),
\end{gather}
where $\alpha$ is the CWG coupling constant, $F=dA$ is the strength of the Maxwell field, and $C_{\mu \nu \rho \sigma}$ is the canonical Weyl tensor \cite{LiuLu2013}. Varying this action gives the Maxwell equation and the fourth-order gravitational field equation
\begin{equation}\label{EOM}
\begin{split}
&\nabla^{\mu}F_{\mu\nu}=0,\\
&\quad -\alpha(2\nabla^{\rho}\nabla^{\sigma}+R^{\rho\sigma})C_{\mu\rho\sigma\nu}+\frac{2}{3}\alpha(F_{\mu\nu}-\frac{1}{4}F^2g_{\mu\nu})=0,
\end{split}
\end{equation}
which results in the following exact black hole solution in Boyer-Lindquist coordinates \cite{MannheimKazanas1991,LiuLu2013, Gartlan:2026wdz}\footnote{The form of the metric in \eqref{metric} differs from that found in Ref. \cite{LiuLu2013} by the temporal diffeomorphism $t\rightarrow \frac{t}{\Xi}$.}. 
\begin{equation}\label{metric}
\begin{split}
ds^2=&\rho^2\left(\frac{dr^2}{\Delta_r}+\frac{d\theta^2}{\Delta_{\theta}}\right)
+ \frac{\Delta_{\theta}\sin^2\theta}{\rho^2\Xi^2}
\left(a\,dt-(r^2+a^2)d\phi\right)^2 \\
&-\frac{\Delta_r}{\rho^2\Xi^2}
\left(dt-a\sin^2\theta d\phi\right)^2, \\
A=&-\frac{qr}{\rho^2\Xi}\left(dt-a\sin^2\theta d\phi\right),
\end{split}
\end{equation}
with
\begin{equation}\label{metric functions}
\begin{split}
& \rho^2=r^2+a^2\cos^2\theta,\quad
\Delta_{\theta}=1+\frac{1}{3}\Lambda a^2 \cos^2\theta,\quad
\Xi=1+\frac{1}{3}\Lambda a^2,\\
& \Delta_r=\left(r^2+a^2\right)\left(1-\frac{1}{3}\Lambda r^2\right)-2Mr+\frac{q^2 r^3}{6M}.
\end{split}
\end{equation}
Here $M$, $q$, $a$, and $\Lambda$ denote the mass, charge, rotation parameter, and cosmological constant. The case $q=0$ reduces to the Kerr-(A)dS form, while the charged case is not the Kerr-Newman-(A)dS geometry. The important difference is the CWG charge contribution $q^2r^3/(6M)$ in $\Delta_r$, whose radial dependence changes the horizon structure and later enters the photon effective potentials. We focus on $\Lambda>0$, so the spacetime is asymptotically de Sitter. In geometrized units, \(G=c=1\), the parameters \(M\), \(a\), and \(r\) have dimensions of length, while the cosmological constant satisfies \([\Lambda]=L^{-2}\). Unlike the Kerr-Newman-(A)dS geometry, in which the corresponding charge parameter has dimensions \([q]=L\) and enters \(\Delta_r\) quadratically as \(q^2\), the CWG metric contains the charge contribution \(q^2r^3/(6M)\). Since \(\Delta_r\) must have dimensions of \(L^2\), dimensional consistency requires
\[
\left[\frac{q^2r^3}{M}\right]=L^2
\quad\Longrightarrow\quad [q]=1.
\]
Thus, in the CWG metric, the corresponding dimensionless parameters are \(a/M\), \(r/M\), \(q\), and \(\Lambda M^2\). We set $M=1$ throughout this paper.

The horizons are located at the real positive roots of $\Delta_r=0$. With $M=1$ this condition is
\begin{gather}
\Delta_r=(r^2+a^2)\left(1-\frac{1}{3}\Lambda r^2\right)-2r+\frac{q^2r^3}{6}=0.
\end{gather}
For parameter values describing a de Sitter black hole, $\Delta_r$ has three positive roots. We denote them, in increasing order, by $r_-$, $r_h$, and $r_c$, corresponding to the Cauchy horizon, the outer event horizon, and the cosmological horizon. A fourth root is negative for the parameter ranges considered here and is not physically relevant for the exterior domain.

The boundaries of the black hole parameter region occur when adjacent horizons merge. We therefore locate extremal curves by solving the double-root conditions
\begin{equation}
\Delta_r(r_\ast)=0,\qquad \Delta_r'(r_\ast)=0,
\end{equation}
where $r_\ast=r_-=r_h$ gives the inner/outer horizon merger and $r_\ast=r_h=r_c$ gives the event/cosmological horizon merger. We compare these curves with the corresponding Kerr-Newman-de Sitter ($KNdS$) curves to isolate the effect of the CWG charge term.

\begin{figure*}
	\centering
\includegraphics[width=1\linewidth, height=6cm]{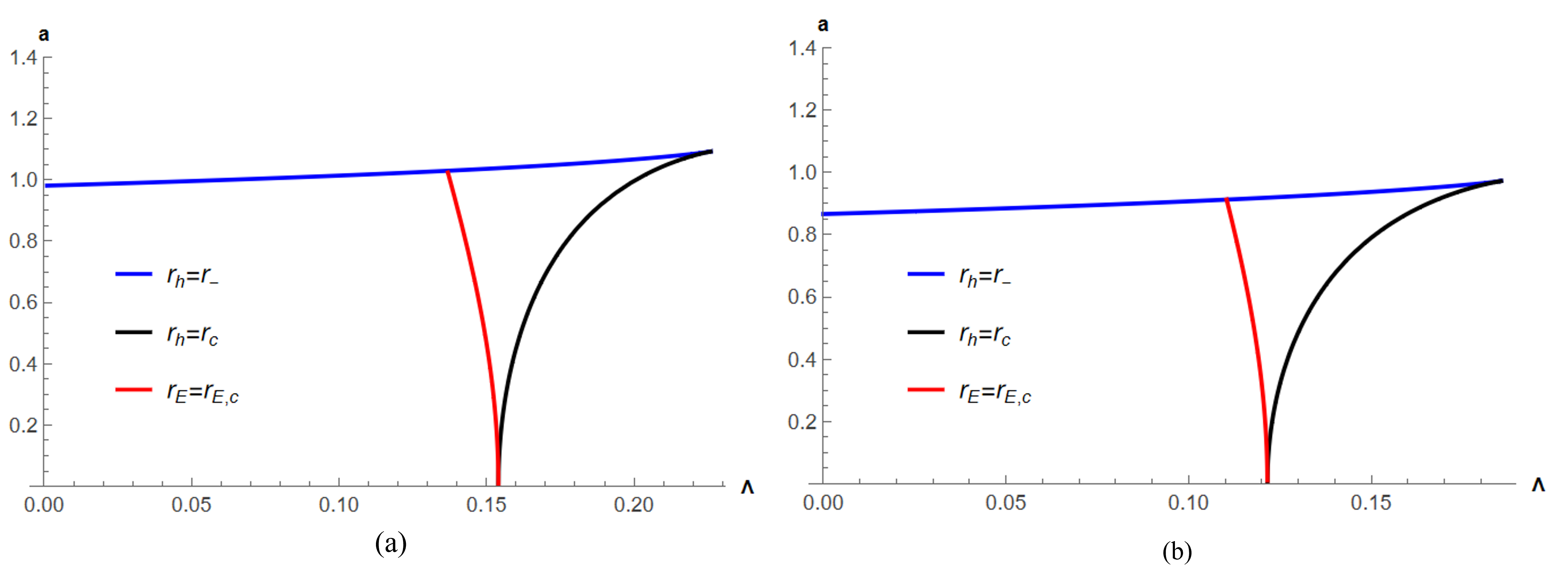}
\caption{Parameter space $(a,\Lambda)$ for the charged rotating de Sitter black hole in CWG, labeled $KNdSCG$ in panel (a), and for the corresponding $KNdS$ spacetime in panel (b). The blue curve denotes the inner/outer horizon merger $r_h=r_-$, while the black curve denotes the event/cosmological horizon merger $r_h=r_c$. The black hole region lies between these two horizon-extremality curves. The red curve marks the merger of the black hole and cosmological stationary-limit surfaces, $r_E=r_{E,c}$, and indicates where the black hole ergoregion ceases to exist. We set $q=0.5$.}
\label{ExtCond}
\end{figure*}

\begin{figure*}
	\centering
\includegraphics[width=1\linewidth, height=6cm]{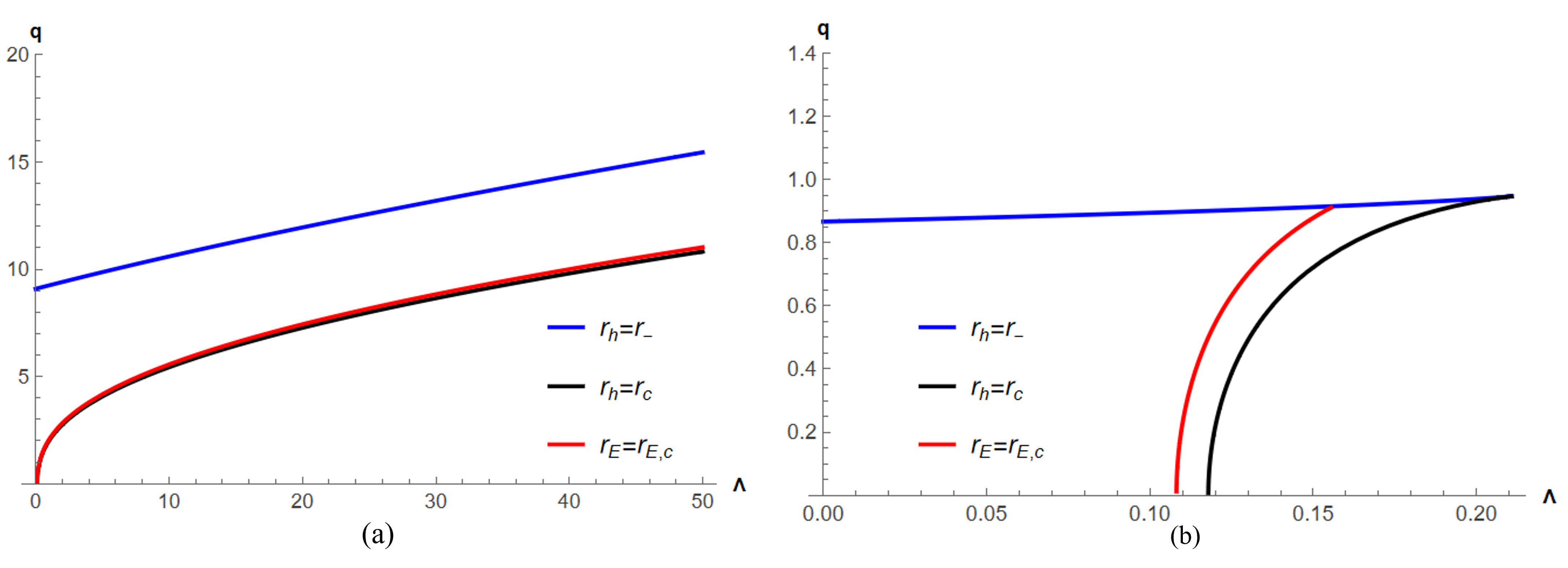}
\caption{Parameter space $(q,\Lambda)$ for the charged rotating de Sitter black hole in CWG, labeled $KNdSCG$ in panel (a), and for the corresponding KNdS spacetime in panel (b). The blue and black curves again denote the horizon-extremality conditions $r_h=r_-$ and $r_h=r_c$, respectively. The red curve marks $r_E=r_{E,c}$, the merger of the black hole and cosmological stationary-limit surfaces. We set $a=0.5$.}
\label{ExtCond2}
\end{figure*}

The stationary-limit surfaces are defined by $g_{tt}=0$, or equivalently
\begin{equation}
\Delta_r-a^2\Delta_\theta\sin^2\theta=0 .
\end{equation}
The red curves in Figs.~\ref{ExtCond} and \ref{ExtCond2} mark where the black hole and cosmological stationary-limit surfaces meet. These curves diagnose the existence of a black hole ergoregion; they are not horizon extremality curves.
Figure~\ref{ExtCond} shows that the allowed region in the $(a,\Lambda)$ plane has the same qualitative shape in $KNdSCG$ and in $KNdS$. For fixed charge, the $KNdSCG$ curve permits a slightly larger range of spin and cosmological constant before the event horizon merges with either the Cauchy horizon or the cosmological horizon.
The difference between $KNdSCG$ and $KNdS$ is more pronounced in the $(q,\Lambda)$ plane, as shown in Fig.~\ref{ExtCond2}. In $KNdS$, charge is bounded in the familiar way by horizon extremality. In the $KNdSCG$ geometry, the cubic charge term changes the large-$r$ behavior of $\Delta_r$, so the allowed region extends to much larger values of $q$ over the plotted range. This feature is a geometric consequence of the CWG radial function and should not be interpreted as an astrophysical charge estimate.
\begin{figure*}
	\centering
\includegraphics[width=1\textwidth]{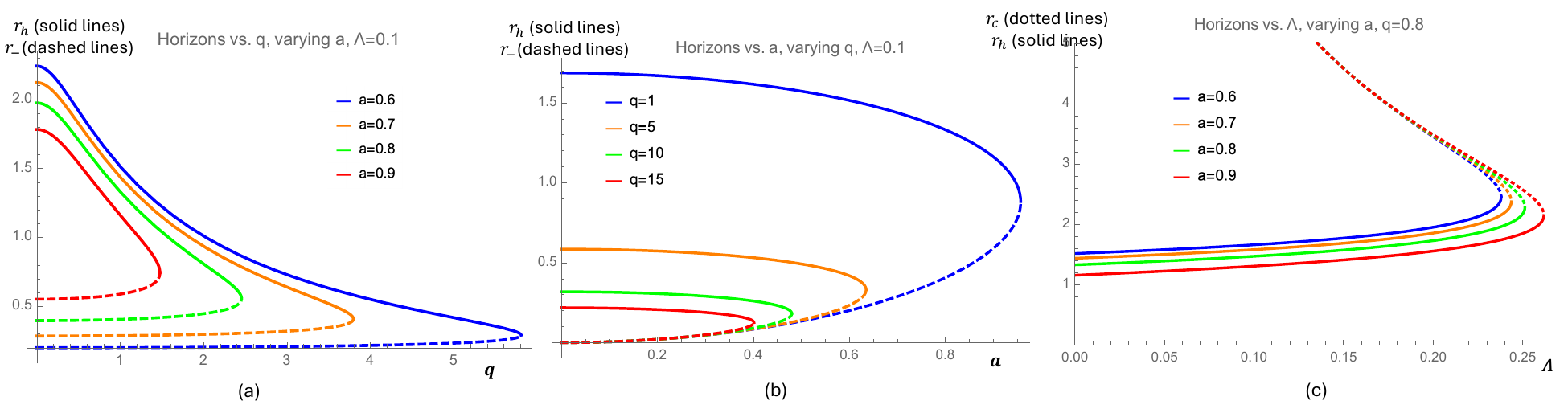}
\caption{Horizon radii for the $KNdSCG$ black hole. Panel (a) shows the Cauchy horizon $r_-$ and event horizon $r_h$ as functions of charge $q$ for several values of spin $a$ at fixed $\Lambda=0.1$. Panel (b) shows $r_-$ and $r_h$ as functions of spin $a$ for several values of charge $q$ at fixed $\Lambda=0.1$. Panel (c) shows the event horizon $r_h$ and cosmological horizon $r_c$ as functions of $\Lambda$ for several values of spin at fixed $q=0.8$. The value $\Lambda=0.1$ is chosen for visual clarity in these dimensionless plots. }
\label{horizons}
\end{figure*}
Figure~\ref{horizons} gives a direct view of how the horizon radii move inside the allowed parameter region. Increasing $q$ decreases the event-horizon radius and increases the Cauchy-horizon radius until the two merge at the charge extremal value. Increasing $a$ produces the same qualitative merger pattern, although the location of the merger depends strongly on $q$. As $\Lambda$ increases, the cosmological horizon moves inward while the event horizon moves outward, and the two meet at event/cosmological horizon merger boundary. For very small positive $\Lambda$, including astrophysically relevant values, the cosmological horizon is far from the black hole and the local photon dynamics near $r_h$ are dominated by the mass, spin, charge, and plasma terms.

The necessary condition on the parameter space for the existence of an outer event horizon, $r_h$, can be analytically obtained. Note first that extremality will occur, by definition, when the metric function $\Delta_r$ has a double root. Importing basic knowledge from calculus, candidate extremal values $r_H$ will therefore ocuur when
\begin{equation}
\Delta_r(r_H)=0 \quad \text{and} \quad \Delta_r'(r_H)=0.
\end{equation}
The second condition, that the derivative of $\Delta_r$ be zero at the degenerate values, yields a cubic equation in $r_H$
\begin{equation}
-\frac{4\Lambda}{3}r_H^3+\frac{q^2}{2}r_H^2+2\left(1-\frac{\Lambda a^2}{3}\right)r_H-2=0.
\end{equation}
Using the cubic formula, we can obtain the three solutions to the above cubic in terms of the black hole parameters. Picking the positive solution corresponding to the $r_h=r_-$ extremality, we see that
\begin{equation}
r_h=r_-=\frac{2}{q^2}\left(\sqrt{1+q^2}-1\right)+\mathcal{O}(\Lambda)
\end{equation}
to leading order in small $\Lambda$. Plugging our full expression for extremal $r_H$ when $r_h=r_-$ into the horizon equation $\Delta_r(r_H)=0$, we obtain the extremality condition
\begin{equation}
\begin{split}
0={}&a^2+\frac{4}{3q^4}\Bigl[2\sqrt{1+q^2}-2\\
&\qquad-\left(3-2\sqrt{1+q^2}\right)q^2\Bigr]\\
&+\frac{64\Lambda}{3q^8}\Bigl[\left(\sqrt{1+q^2}-2\right)q^2\\
&\qquad+2\left(\sqrt{1+q^2}-1\right)\Bigr]\\
&-\frac{4\Lambda}{3q^4}\Bigl[a^2\left(q^2-2\sqrt{1+q^2}+2\right)+4\Bigr]\\
&+\mathcal{O}\left(\Lambda^2\right).
\end{split}
\end{equation}
It follows that the condition forbidding a naked singularity is given by
\begin{equation}
\begin{split}
a^2\leq{}&\frac{4}{q^4}\Bigg[ \frac{\left(2\sqrt{1+q^2}-3\right)q^6}{3q^4-4\Lambda q^2+8\Lambda\left(\sqrt{1+q^2}-1\right)}\\
&+\frac{2q^4\left(2\Lambda+\sqrt{1+q^2}-1\right)}{3q^4-4\Lambda q^2+8\Lambda\left(\sqrt{1+q^2}-1\right)}\\
&-\frac{16\Lambda\left(\sqrt{1+q^2}-2\right)q^2}{3q^4-4\Lambda q^2+8\Lambda\left(\sqrt{1+q^2}-1\right)}\\
&-\frac{32\Lambda\left(\sqrt{1+q^2}-1\right)}{3q^4-4\Lambda q^2+8\Lambda\left(\sqrt{1+q^2}-1\right)}\Bigg]\\
&+\mathcal{O}\left(\Lambda^2\right).
\end{split}
\end{equation}
For $\Lambda=0$, this can be simplified to
\begin{equation}
a^2 \leq \frac{8}{3q^4} \left[\left(1+q^2\right)^{3/2}-1\right]-\frac{4}{q^2}.
\end{equation}
Observe that taking the limit of both sides of the above inequality as $q\to0$ yields the Kerr extremality condition
\begin{equation}
a^2\leq 1,
\end{equation}
reflecting the fact that the KNdSCG spacetime reduces to the Kerr solution for $q=\Lambda=0$. Away from the $q=0$ regime, we see that the allowed range of $a^2$ varies significantly from the GR counterpart of $a^2\leq 1-q^2$ for the Kerr-Newman spacetime. In contrast to Kerr-Newman spacetime, where extremality defines a compact region in $(a,q)$-space, the KNdSCG solution for $\Lambda=0$ exhibits a non-compact extremal locus in which the charge parameter does not have an induced upper bound, but instead continuously deforms the allowed spin bound for all $q\in \R$. For the Kerr-Newman spacetime, $q^2\leq 1$ must be obeyed to be consistent with positive $a^2$ and the GR horizon degeneracy relation $a^2\leq 1-q^2$. In contrast, notice that
\begin{equation}
\frac{8}{3q^4} \left[\left(1+q^2\right)^{3/2}-1\right]-\frac{4}{q^2} > 0, \text{ for all $q\in\R$.}
\end{equation}
Hence, cosmic censorship via horizon degeneracy does not impose an upper bound on the charge parameter of the KNdSCG spacetime for $\Lambda=0$ (hereafter KNCG). Instead, the KNCG black hole solution remains well-defined for arbitrarily large charge, in the sense that the extremality condition admits horizon solutions for all $q\in \R$. Indeed, the upper bound imposed on $a^2$ by horizon degeneracy decays as $q\to\infty$, approaching zero from above reflecting that highly charged KNCG black holes are forced toward non-rotating (or very slowly rotating) configurations.
\section{Plasma Inclusion in Hamilton-Jacobi Formulism} 
The photon dynamics near a black hole is not only affected by the local curves geometry, but any optical mediums the black hole is immersed in. One such medium is plasma, and to model this we now include a cold, pressureless, non-magnetized plasma as a dispersive optical medium in our ray-optics. The plasma is treated as a test medium, so it changes photon propagation but does not back-react on the CWG black hole geometry. In geometric optics, photon rays in plasma are generated by the Hamiltonian \cite{PerlickTsupkoBisnovatyiKogan2015,Perlick2017PlasmaKerr,ChowdhuriBhattacharyya2021PlasmaShadow}
\begin{equation}\label{eq:PlasmaHamiltonian}
\mathcal{H}=\frac{1}{2}\left(g^{\mu\nu}p_\mu p_\nu+\omega_p^2(x)\right)=0,
\end{equation}
where $\omega_p^2(x)=e^2 n_e(x)/m_e$ is the local plasma frequency at some spacetime coordinates $x$. The equivalent refractive index is
\begin{equation}\label{eq:IndexofRefraction}
n^2(x,\omega)=1-\frac{\omega_p^2(x)}{\omega^2(x)},
\end{equation}
where $\omega(x)$ is the photon frequency measured by a local observer and changes along the ray because of gravitational redshift \cite{Rogers2015}. Propagating rays require $\omega(x)>\omega_p(x)$ since when $\omega(x)<\omega_p(x)$, the refractive index becomes imaginary and the geometric-optics ray description has no propagating solution. We therefore restrict the plasma strength so that the ratio $\omega_p^2/\omega^2$ remains below unity along the relevant photon trajectories.

For a rotating spacetime, the Hamilton-Jacobi equation remains separable only for restricted plasma profiles. Following the Kerr-plasma separability condition \cite{Perlick2017PlasmaKerr}, we impose
\begin{equation}\label{eq:SeperablePlasFreqForm}
\rho^2\omega_p^2(r,\theta)=f_r(r)+f_\theta(\theta).
\end{equation}
After dividing the separated equations by the conserved photon energy squared, we use the same symbols $f_r$ and $f_\theta$ for the corresponding dimensionless plasma functions. In the numerical examples below we take
\begin{equation}
\begin{aligned}
f_r(r)&=\left(\frac{\omega_c}{\omega_0}\right)^2\sqrt{M^3r},\\
f_\theta(\theta)&=\left(\frac{\omega_c}{\omega_0}\right)^2M^2\left(1+2\sin^2\theta\right),
\end{aligned}
\end{equation}
where $\omega_0$ sets the photon-frequency scale and $\omega_c$ is a scaling factor controls the plasma density, or "strength". This profile contains both radial and angular structure, allowing the plasma to affect the radial photon region and the projected shadow shape.
\begin{figure*}
\centering
\includegraphics[width=1\textwidth]{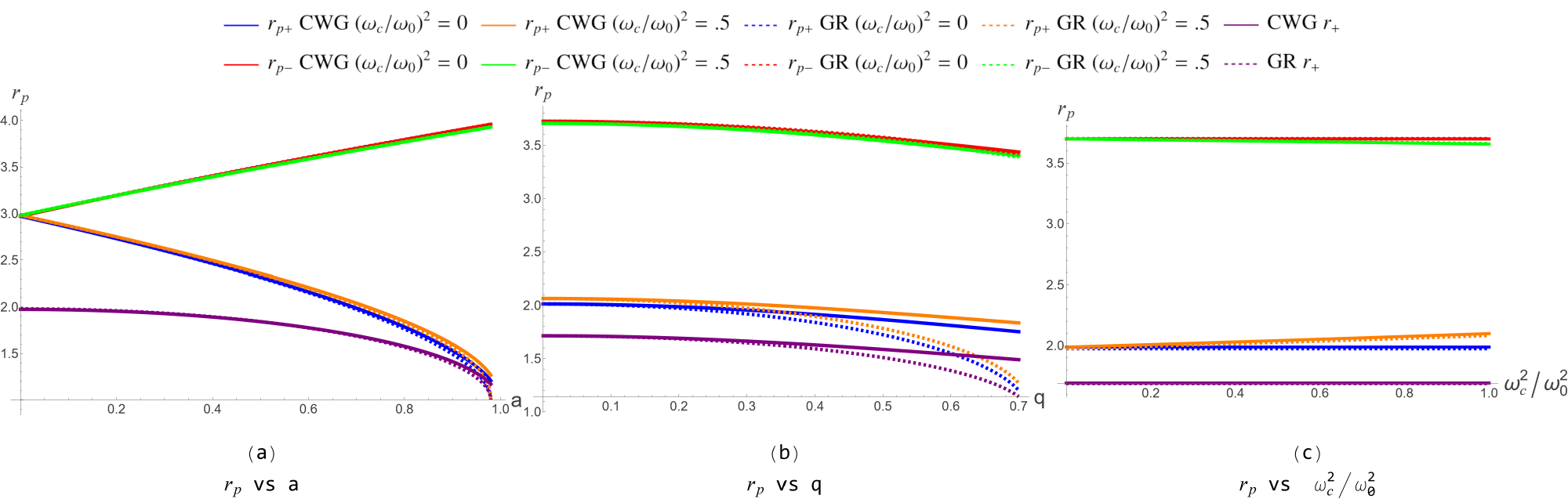}
\caption{Equatorial circular photon-orbit radii in CWG and GR. Solid curves denote CWG, dashed curves denote GR, and purple curves show the corresponding outer event horizon $r_h$. The lower branch $r_{p+}$ is the co-rotating orbit and the upper branch $r_{p-}$ is the counter-rotating orbit. Panel (a) varies spin $a$ at fixed $q=0.2$; panel (b) varies charge $q$ at fixed $a=0.7$; panel (c) varies the plasma strength $(\omega_c/\omega_0)^2$ at fixed $a=0.7$ and $q=0.2$. Unless varied, $(\omega_c/\omega_0)^2=0.5$ and $\Lambda=10^{-26}$.}
\label{fig:rp}
\end{figure*} 
\subsection{Photon trajectories in plasma}\label{sect:Geodesics}
For the Hamilton-Jacobi action we use
\begin{equation}
S=-Et+l_\phi\phi+S_r(r)+S_\theta(\theta),
\end{equation}
where $E=-p_t$ and $l_\phi=p_\phi$ are conserved by stationarity and axisymmetry. Substituting the inverse metric associated with Eq.~\eqref{metric} into Eq.~\eqref{eq:PlasmaHamiltonian}, multiplying by $\rho^2$, and using Eq.~\eqref{eq:SeperablePlasFreqForm} gives the compact separated equation
\begin{equation}
\begin{aligned}
&\Delta_r\left(\frac{dS_r}{dr}\right)^2
+\Delta_\theta\left(\frac{dS_\theta}{d\theta}\right)^2
-\frac{\Xi^2\left[E(r^2+a^2)-a l_\phi\right]^2}{\Delta_r} \\
&\quad
+\frac{\Xi^2\left(l_\phi\csc\theta-aE\sin\theta\right)^2}{\Delta_\theta}
+f_r(r)+f_\theta(\theta)=0 .
\label{eq:jacobi}
\end{aligned}
\end{equation}
Introducing the separation Carter constant $\kappa$, this equation becomes
\begin{equation}
\left\{
\begin{aligned}
\Delta_r\left(\frac{dS_r}{dr}\right)^2
&=\frac{\Xi^2\left[E(r^2+a^2)-a l_\phi\right]^2}{\Delta_r}
-\kappa-f_r(r),\\
\Delta_\theta\left(\frac{dS_\theta}{d\theta}\right)^2
&=\kappa-f_\theta(\theta)
-\frac{\Xi^2\left(l_\phi\csc\theta-aE\sin\theta\right)^2}{\Delta_\theta}.
\end{aligned}
\right.
\end{equation}
The canonical relations $p_r=\partial S_r/\partial r=(\rho^2/\Delta_r)\dot r$ and $p_\theta=\partial S_\theta/\partial\theta=(\rho^2/\Delta_\theta)\dot\theta$ then give
\begin{equation}
\begin{aligned}
\left(\frac{\rho^2}{E}\right)^2\dot r^2
=R(r)
&=\Xi^2\left(r^2+a^2-aL_E\right)^2\\
&\quad-\Delta_r\left(K_E+f_r(r)\right),
\label{eq:R(r)}
\end{aligned}
\end{equation}
\begin{equation}
\begin{aligned}
\left(\frac{\rho^2}{E}\right)^2\dot\theta^2
=\Theta(\theta)
&=\Delta_\theta\left(K_E-f_\theta(\theta)\right)\\
&\quad-\Xi^2\left(L_E\csc\theta-a\sin\theta\right)^2,
\label{eq:Th(th)}
\end{aligned}
\end{equation}
where
\begin{equation}
L_E=\frac{l_\phi}{E},\qquad K_E=\frac{\kappa}{E^2}.
\end{equation}
Equations~\eqref{eq:R(r)} and \eqref{eq:Th(th)} are the main dynamical input for the rest of the paper. The difference between CWG and GR enters through $\Delta_r$, while the plasma enters through $f_r$ and $f_\theta$. Their product in Eq.~\eqref{eq:R(r)} is important: the same plasma profile produces different photon dynamics in CWG and GR because it is weighted by a different radial metric function.
\subsection{Circular photon orbits in the equatorial plane}
Circular photon orbits occur at radii $r_p$ where the radial potential has a double root,
\begin{equation}
R(r_p)=0,\qquad R'(r_p)=0.
\end{equation}
For the equatorial orbits considered here we additionally impose $\Theta(\pi/2)=0$; Because both the metric and our choice of $f_\theta(\theta)$ are invariant under $\theta\to\pi-\theta$, photons launched in the equatorial plane with $\dot{\theta}=0$ remain confined to that plane. Defining
\begin{equation}
A(r_p)=r_p^2+a^2-aL_E,
\end{equation}
and using these two equations to eliminate $K_E$, we obtain
\begin{equation}
\Delta_r'(r_p)A^2(r_p)-4r_p\Delta_r(r_p)A(r_p)
+\frac{\Delta_r^2(r_p)}{\Xi^2}f_r'(r_p)=0.
\end{equation}
Thus the two branches are
\begin{equation}
A_\pm(r_p)=\frac{\Delta_r(r_p)}{\Delta_r'(r_p)}
\left[2r_p\pm\sqrt{4r_p^2-\frac{\Delta_r'(r_p)f_r'(r_p)}{\Xi^2}}\right],
\end{equation}
\begin{align}
L_{E,\pm}(r_p)
&=\frac{r_p^2+a^2-A_\pm(r_p)}{a},\\
K_{E,\pm}(r_p)
&=\frac{\Xi^2A_\pm^2(r_p)}{\Delta_r(r_p)}-f_r(r_p).
\end{align}
The allowed photon-orbit radii are obtained numerically by combining these expressions with the angular condition $\Theta(\pi/2)=0$. The two solutions correspond to the co-rotating branch $r_{p+}$ and the counter-rotating branch $r_{p-}$.

Figure~\ref{fig:rp} compares the CWG and GR photon-orbit structure in vacuum and in plasma. Panel (a) shows the usual spin splitting: increasing $a$ pulls the co-rotating orbit inward and pushes the counter-rotating orbit outward. The CWG and GR curves follow the same qualitative trend, but the co-rotating CWG branch remains slightly larger than its GR counterpart over much of the high-spin range. Panel (b) shows a stronger theory dependence. As $q$ increases, the GR branches move more rapidly toward their extremal boundary, while the CWG branches vary more gradually because the charge enters $\Delta_r$ through the cubic term $q^2r^3/(6M)$. Panel (c) isolates the plasma effect. Increasing $(\omega_c/\omega_0)^2$ shifts the co-rotating CWG orbit outward more visibly than the corresponding GR orbit, whereas the counter-rotating branch is nearly insensitive to the same plasma change. The important take-away here is that plasma is not merely an additive environmental correction; its influence is filtered through the background geometry, so the same plasma model leaves different signatures in CWG and in GR.
\section{Plasma Influence on Deflection Angle}
In this section, we evaluate the deflection angle of photons with a given impact parameter ($L_E$) and analyze how the plasma profile influences the deflection angle for both theories. The deflection is a measure of gravitational lensing produced by same radial potential that determines the photon region and the shadow. We utilized the methods for computing finite distance deflection angle described in \cite{GrenzebachPerlickLaemmerzahl2014}. The deflection angle, $\alpha$, is given by
\[
\alpha = \Psi_O - \Psi_S + \phi_{\text{OS}},
\]
where $\Psi_O$ and $\Psi_S$ are the local angles between the photon direction and the outward radial direction at the observer and the source position, respectively and $\phi_{\text{OS}}$ is the azimuthal separation angle along the ray. This definition remains meaningful when both endpoints are at finite radii, as required for the de Sitter geometries considered here.
\begin{figure}[h]
    \centering
    \includegraphics[width=0.9\linewidth]{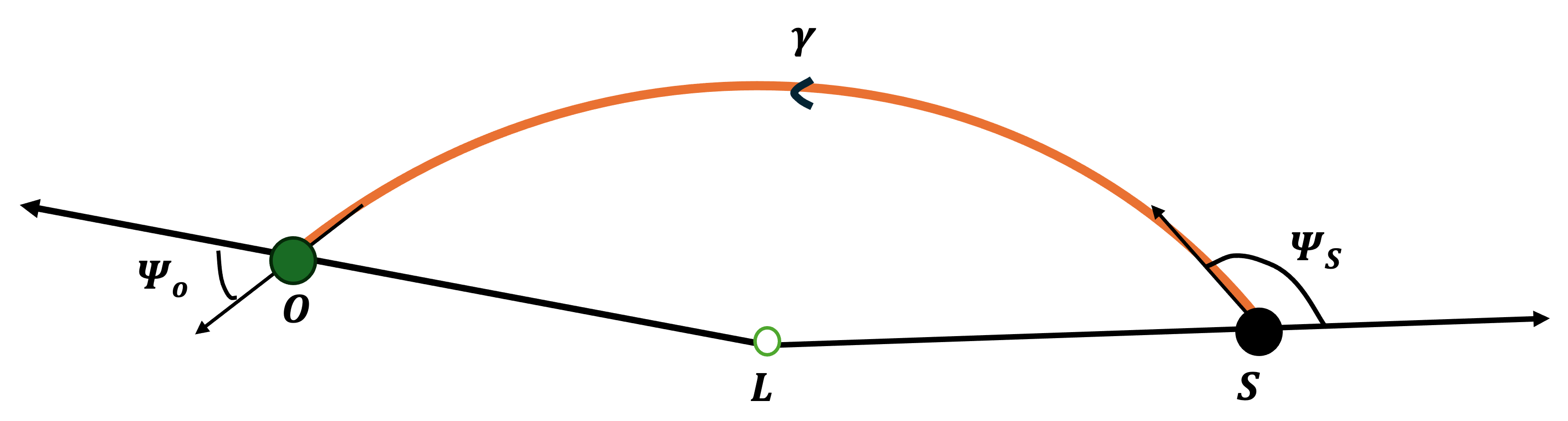}
    \caption{Schematic illustration of the finite-distance deflection geometry. Here, $L$ denotes the lens, and $\gamma$ represents the light trajectory from the source $S$ to the observer $O$. The angles $\Psi_S$ and $\Psi_O$ are measured between the outward radial direction and the tangent to $\gamma$ at the source and observer, respectively.}
    \label{fig:GB}
\end{figure} 
We evaluate the endpoint angles in local zero-angular-momentum-observer (ZAMO) frame at source radius $r_S$ and observer radius $r_O$. As shown Fig.~\ref{fig:GB}, to retain the correct quadrant for the ingoing and outgoing portions of the ray, we write the angles $\Psi_O$ and $\Psi_S$
\begin{equation}
\Psi_O := \left. \tan^{-1}\left(\frac{p^{(\phi)}}{p^{(r)}}\right) \right|_{r=r_O}, \, \, \, \Psi_S := \left. \tan^{-1}\left(\frac{p^{(\phi)}}{p^{(r)}}\right) \right|_{r=r_S},
\end{equation}
where $p^{(\phi)}\equiv \frac{p_{\phi}}{\sqrt{g_{\phi \phi}}}$ and $p^{(r)}\equiv \frac{p_r}{\sqrt{g_{rr}}}$ are the local azimuthal and radial four momenta of the photon in the ZAMO frame.
\begin{figure*}
\centering
\includegraphics[width=1\textwidth]{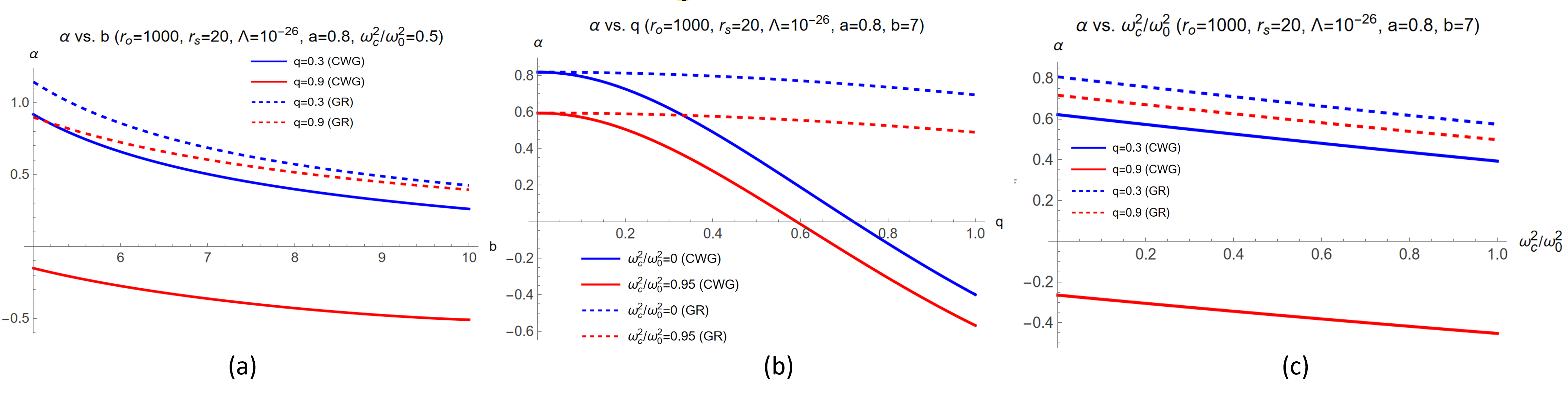}
\caption{Signed finite-distance deflection angle for CWG (solid curves) and GR (dashed curves). Panel (a) varies the impact parameter $L_E$ at $\omega_c^2/\omega_0^2=0.5$ for $q=0.3$ and $q=0.9$. Panel (b) varies $q$ at $L_E=7$ in vacuum and at $\omega_c^2/\omega_0^2=0.95$. Panel (c) varies the plasma strength at $L_E=7$ for $q=0.3$ and $q=0.9$. In all panels, $M=1$, $a=0.8$, $\theta=\pi/2$, $\Lambda=10^{-26}$, $r_O=1000$, and $r_S=20$.}
\label{fig: deflectangle}
\end{figure*}
For an equatorial ray, $\theta=\pi/2$ and $\dot\theta=0$. The angular potential must therefore satisfy $\Theta(\pi/2)=0$. For the reflection-symmetric plasma profile used here, $\Theta'(\pi/2)=0$ automatically, and Eq.~\eqref{eq:Th(th)} gives
\begin{equation}
K_E=f_\theta\!\left(\frac{\pi}{2}\right)+\Xi^2(L_E-a)^2.
\label{eq:equatorialKE}
\end{equation}
Solving $E=-g_{t\mu}\dot{x}^{\mu}$ and $l_\phi=g_{\phi\mu}\dot{x}^{\mu}$ for the azimuthal velocity yields
\begin{equation}
\dot{\phi} = \frac{ \Xi^2 E}{r^2 \Delta_r} [a(r^2+a^2-\Delta_r)+L_E(\Delta_r-a^2)].
\end{equation}
Then, using \eqref{eq:R(r)} for $\dot{r}^2$, we get that
\begin{equation}
\left(\frac{d\phi}{dr}\right)^2=
\frac{\Xi^4\left[a\left(r^2+a^2-\Delta_r\right)+L_E\left(\Delta_r-a^2\right)\right]^2}
{\Delta_r^2\left\{\Xi^2\left(r^2+a^2-aL_E\right)^2-\Delta_r\left[K_E+f_r(r)\right]\right\}}.
\label{eq:dphidrdeflection}
\end{equation}
The accumulated azimuthal separation $\phi_{\text{OS}}$ is given by
\begin{equation}
\phi_{\text{OS}}= \int_{R_c}^{r_S} \abs{\frac{d\phi}{dr}}dr + \int_{R_c}^{r_O} \abs{\frac{d\phi}{dr}}dr,
\end{equation}
where $R_c$ is the radial turning point determined by $\dot{r}(R_c) = 0$). We use the CWG radial function in the conformal-gravity calculation and the KNdS radial function in the GR calculation, while keeping the endpoint geometry, plasma profile, and conserved quantities the same.

Figure~\ref{fig: deflectangle} compares the finite deflection angle $\alpha$ predicted in CWG (KNdSCG) with the corresponding GR result (KNdS). It shows that every plotted curve slopes downward as the impact parameter $L_E$, black hole charge $q$, or plasma frequency $\omega_c^2/\omega_0^2$ is increased, but the physical interpretation depends on the sign of $\alpha$. In panel (a), the positive CWG and GR branches approach smaller positive angles as $L_E$ grows. The high-charge CWG branch is already negative and moves to more negative values, so its absolute value increases rather than decreases. 

Panel (b) illustrates the dependence of the deflection angle on the charge parameter $q$ for $L_E=7$, which shows the strongest theory dependence. At fixed impact parameter $L_E=7$, the GR curves decreases slowly with $q$ and remain positive, while the CWG curves fall much more rapidly and cross $\alpha=0$. The plasma curve lies below the vacuum curve in each theory, but the plasma-induced shift is amplified in CWG. 

Panel (c) shows the dependence of $\alpha$ on the normalized plasma frequency. For both values of $q$, the CWG deflection angle decreases approximately monotonically as $\omega_c^2/\omega_0^2$ increases. The effect is particularly pronounced for $q=0.9$, for which the deflection angle is already negative and becomes increasingly negative as the plasma contribution increases. This also confirms the pattern at fixed charge: stronger plasma lowers the signed angle in both theories, with the largest change occurring for the high-charge CWG case. The GR predictions also decrease with increasing plasma frequency, but much more weakly and remain positive. These results demonstrate that the finite deflection angle in CWG can differ qualitatively from the GR prediction. While increasing $L_E$, $q$, or the plasma frequency generally reduces $\alpha$, the dependence on these parameters is considerably stronger in CWG. Most notably, the CWG deflection angle can change sign for sufficiently large $q$, with plasma effects further enhancing this deviation from the GR behavior.
\section{Black Hole Shadow and Observables}
\subsection{Black hole shadow for a finite observer}
For asymptotically flat rotating black holes, the shadow is often described by Bardeen's celestial coordinates for an observer at spatial infinity \cite{Bardeen1973}. That construction is not appropriate for the charged CWG geometry considered here. When $q\neq0$, the radial function contains the term $q^2r^3/(6M)$, so the metric does not approach the asymptotically flat Kerr form. In addition, for $\Lambda>0$ the exterior observer region is bounded by the cosmological horizon. The shadow must therefore be defined operationally on the local sky of an observer at finite position $(r_o,\theta_o)$. We follow the tetrad projection method used for non-asymptotically flat and plasma-filled black hole spacetimes \cite{GrenzebachPerlickLaemmerzahl2014,Perlick2017PlasmaKerr,PerlickTsupko2022Review,ChowdhuriBhattacharyya2021PlasmaShadow}, and then map the local observation angles to dimensionless screen coordinates.

For compactness in the tetrad expressions below, we define
\begin{equation}
\chi=a\sin^2\theta ,
\end{equation}
so that $r^2+a^2=\rho^2+a\chi$ for $\rho^2 = r^2 + a^2\cos^2\theta$.
An orthonormal coframe adapted to the stationary observer is
\begin{align}
w^{(0)} &= \sqrt{\frac{\Delta_r}{\rho^2}}
\left(\frac{dt}{\Xi} - \frac{a \sin^2\theta}{\Xi} d\phi \right),\\
w^{(1)} &= \sqrt{\frac{\rho^2}{\Delta_\theta}} d\theta,\\
w^{(2)} &= \sqrt{\frac{\Delta_\theta}{\rho^2}} \sin\theta
\left(a \frac{dt}{\Xi} - \frac{\rho^2 + a\chi}{\Xi} d\phi \right),\\
w^{(3)} &= \sqrt{\frac{\rho^2}{\Delta_r}} dr .
\end{align}
The dual tetrad vectors, defined by $w^{(a)}(e_{(b)})=\delta^a_b$, are
\begin{align}
e_{(0)} &= \frac{(\rho^2+a\chi)\Xi\partial_t+a\Xi\partial_\phi}
{\sqrt{\rho^2\Delta_r}},\\
e_{(1)} &= \sqrt{\frac{\Delta_\theta}{\rho^2}}\partial_\theta,\\
e_{(2)} &= -\frac{\chi\Xi\partial_t+\Xi\partial_\phi}
{\sqrt{\rho^2\Delta_\theta}\sin\theta},\\
e_{(3)} &= -\sqrt{\frac{\Delta_r}{\rho^2}}\partial_r .
\end{align}
The sign of $e_{(3)}$ is chosen so that photons arriving from the black hole have positive local radial direction on the observer's sky.
\begin{figure*}
	\centering
    \includegraphics[width=\textwidth]{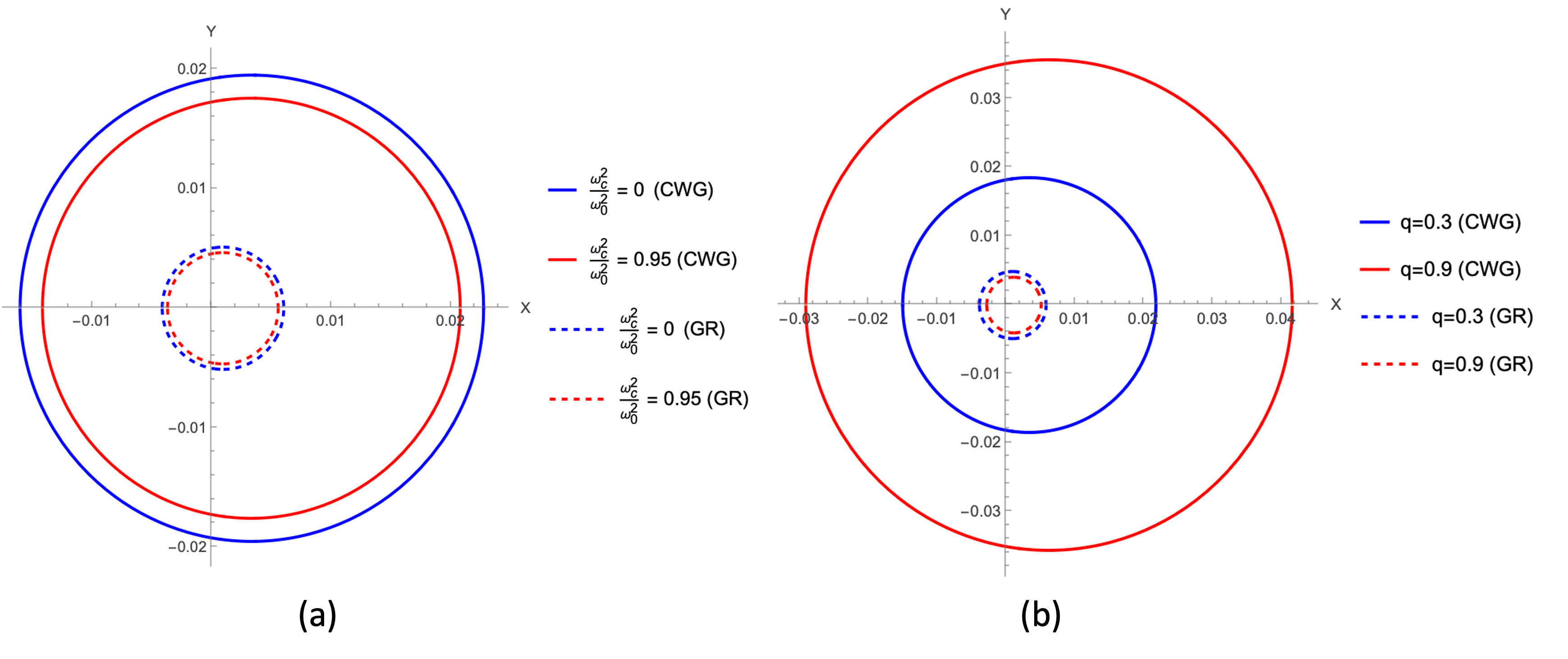}
    \caption{Finite-observer shadow boundaries in CWG and GR. Solid curves denote CWG and dashed curves denote GR. Panel (a) varies the plasma strength, with $\omega_c^2/\omega_0^2=0$ in blue and $\omega_c^2/\omega_0^2=0.95$ in red, at fixed $a=0.3$, $q=0.3$, $r_o=1000$, $\theta_o=\pi/2$, and $\Lambda=10^{-26}$. Panel (b) varies charge, with $q=0.3$ in blue and $q=0.9$ in red, at fixed $a=0.3$, $r_o=1000$, $\theta_o=\pi/2$, $\Lambda=10^{-26}$, and $\omega_c^2/\omega_0^2=0.5$.}
    \label{fig:Shadow-compare}
\end{figure*}

The tetrad converts the photon momentum into quantities measured by the finite observer. Let $p^{(a)}$ denote these local components. The spatial vector $(p^{(1)},p^{(2)},p^{(3)})$ determines the position of the incoming ray on the observer's sky, so we introduce the celestial angles following the finite-observer construction of Refs.~\cite{GrenzebachPerlickLaemmerzahl2014,Perlick2017PlasmaKerr}:
\begin{align}
p^{(1)} &= |{\bf p}|\sin\vartheta\cos\psi,\\
p^{(2)} &= |{\bf p}|\sin\vartheta\sin\psi,\\
p^{(3)} &= |{\bf p}|\cos\vartheta,
\end{align}
where $|{\bf p}|^2=(p^{(1)})^2+(p^{(2)})^2+(p^{(3)})^2$. Here $\vartheta$ is the polar angle measured from the local radial direction and $\psi$ is the azimuthal angle in the screen plane. In vacuum this construction would involve a null local momentum. In plasma, however, the Hamiltonian constraint $\mathcal{H}=0$ gives
\begin{equation}
(p^{(0)})^2-|{\bf p}|^2=\omega_p^2(r_o,\theta_o).
\end{equation}
This is the local form of the plasma dispersion relation: in the orthonormal frame the momentum norm is $-(p^{(0)})^2+|{\bf p}|^2=-\omega_p^2$. The angles are then obtained by taking ratios of the spatial components. The separated angular potential determines the transverse momentum in the $e_{(1)}$--$e_{(2)}$ plane, while the dispersion relation supplies the total spatial momentum. Substituting the constants of motion and the separated potentials from Eqs.~\eqref{eq:R(r)} and \eqref{eq:Th(th)} gives
\begin{align}
\sin\psi
&=
\frac{\left[-\Xi L_E(r_p)+\chi(\theta_o)\right]\csc\theta_o} 
{\sqrt{\Delta_\theta(\theta_o)}\sqrt{K_E(r_p)-f_\theta(\theta_o)}},
\label{eq:sinpsi}\\
\sin\vartheta
&=
\frac{\sqrt{\Delta_r(r_o)}\sqrt{K_E(r_p)-f_\theta(\theta_o)}}
{\sqrt{\left[r_o^2+a^2-a\Xi L_E(r_p)\right]^2
-\Delta_r(r_o)f_p(r_o,\theta_o)}} ,
\label{eq:sinvartheta}
\end{align}
where
\begin{equation}
f_p(r_o,\theta_o)=f_r(r_o)+f_\theta(\theta_o).
\end{equation}
The shadow boundary is generated by spherical photon orbits labeled by $r_p$. The allowed range of $r_p$ is determined by $\sin^2\psi\leq1$, with the two endpoints obtained from $\sin^2\psi=1$. Finally, we project the local sky onto the dimensionless screen coordinates
\begin{align}
X(r_p) &= -2\tan\left(\frac{\vartheta(r_p)}{2}\right)\sin\psi(r_p),\\
Y(r_p) &= -2\tan\left(\frac{\vartheta(r_p)}{2}\right)\cos\psi(r_p).
\end{align}

Figure~\ref{fig:Shadow-compare} shows two complementary effects. In panel (a), increasing the plasma strength decreases the apparent shadow size in both theories. The plasma correction is therefore qualitatively similar in CWG and GR for this parameter choice, but the baseline size remains theory dependent: the CWG shadow is much larger than the GR shadow at the same $a$, $q$, $r_o$, and $\theta_o$. In panel (b), varying charge reveals a sharper distinction between the two theories. Increasing $q$ enlarges the CWG shadow substantially, while the corresponding GR shadow becomes slightly smaller. This opposite charge response follows from the different radial functions: in CWG, the charge enters through the radial term $q^2r^3/(6M)$, whereas in GR the charge contribution has the usual Kerr-Newman form. Thus plasma changes the overall scale of the shadow, but it does not remove the qualitative CWG/GR distinction produced by charge.
\begin{figure}[h]
    \centering
    \includegraphics[width=0.9\linewidth]{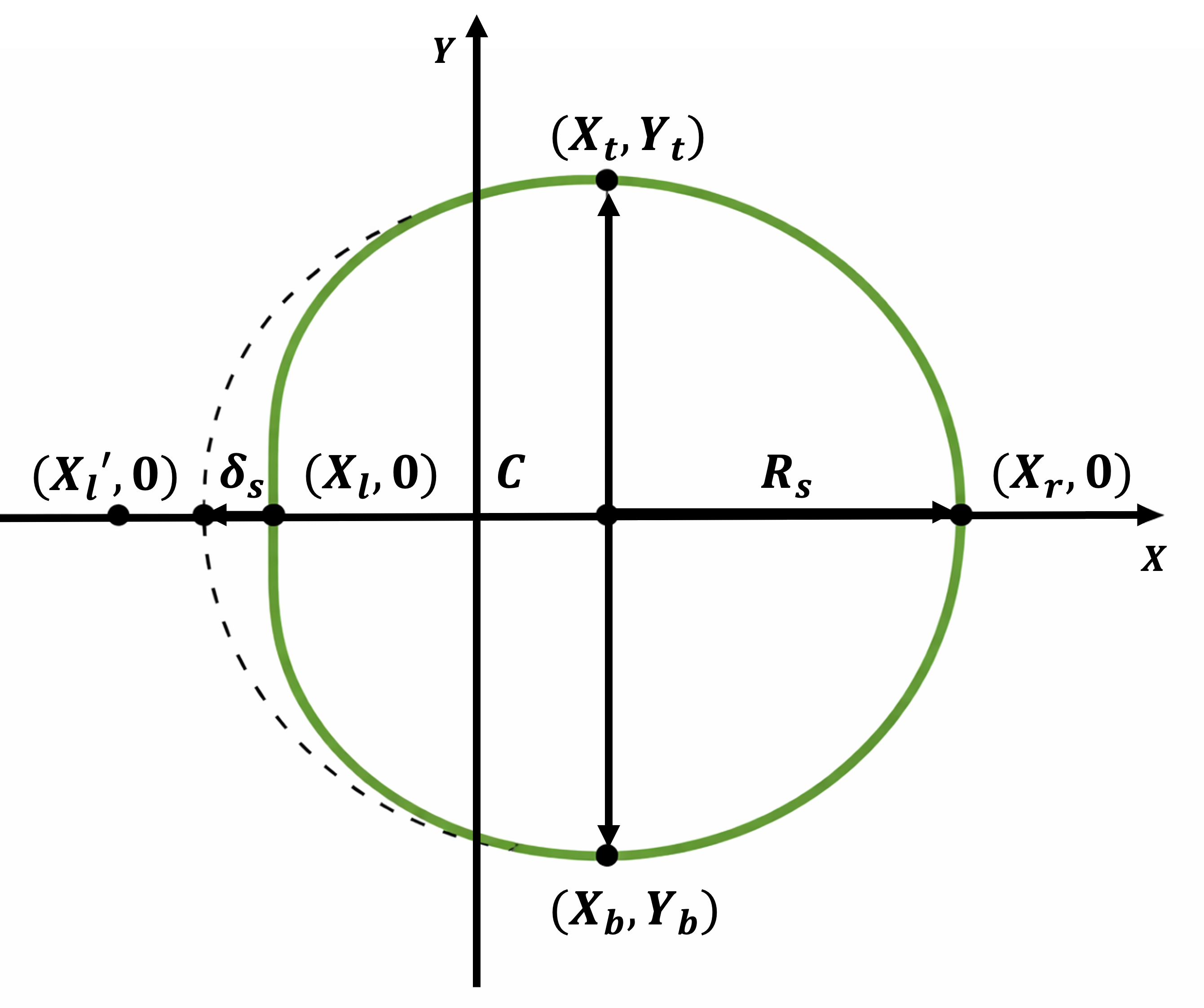}
    \caption{Diagram of the shadow observables with relevant points labeled.}
    \label{fig:ShadowObservDiagram}
\end{figure}

\subsection{Shadow observables}
To summarize the finite-observer shadow boundary, we compute the effective radius and distortion in the dimensionless screen plane. Let $(X_t,Y_t)$ be the top point of the shadow, $(X_r,0)$ the rightmost point, and $(X_l,0)$ the leftmost point, as depicted in Fig.~\ref{fig:ShadowObservDiagram} The reference circle passing through $(X_t,Y_t)$ and $(X_r,0)$ has center $(X_c,0)$, where
\begin{align}
X_c=\frac{1}{2}\left(X_r+X_t+\frac{Y_t^2}{X_t-X_r}\right).
\end{align}
The effective radius and distortion are then
\begin{align}
R_s=\frac{(X_r-X_t)^2+Y_t^2}{2|X_r-X_t|},
\end{align}
\begin{align}
\delta_s=\frac{X_l-(X_c-R_s)}{R_s}.
\end{align}
\begin{figure*}
	\centering
    \includegraphics[width=\textwidth]{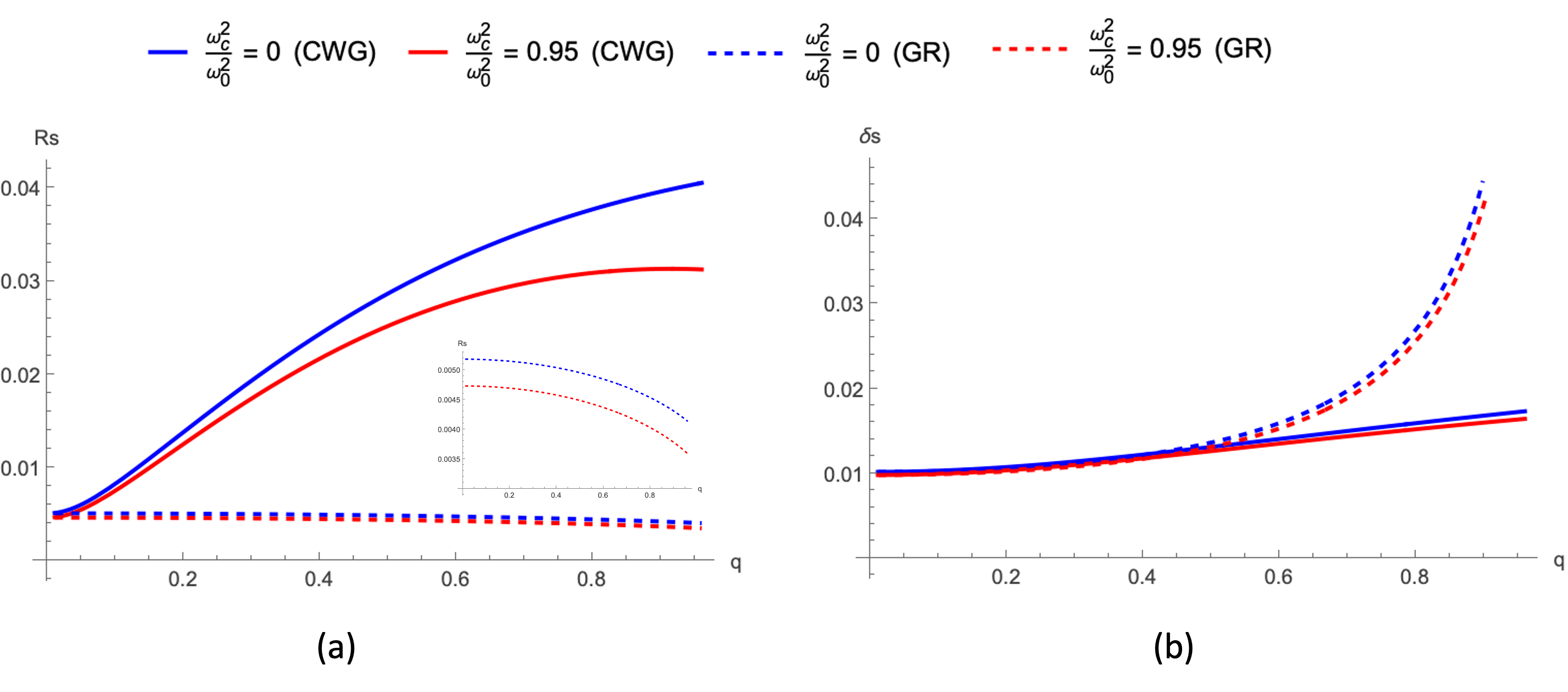}
    \caption{Finite-observer shadow observables as functions of charge $q$. Solid curves denote CWG and dashed curves denote GR. Blue curves show the vacuum case $\omega_c^2/\omega_0^2=0$, while red curves show the plasma case $\omega_c^2/\omega_0^2=0.95$. Panel (a) shows the effective radius $R_s$ The inset provides a magnified view of the dashed curves to highlight their variation with $q$. Panel (b) shows the distortion $\delta_s$. We set $a=0.3$, $r_o=1000$, $\theta_o=\pi/2$, and $\Lambda=10^{-26}$.}
    \label{fig:Observableq}
\end{figure*}
\begin{figure*}
	\centering
    \includegraphics[width=\textwidth]{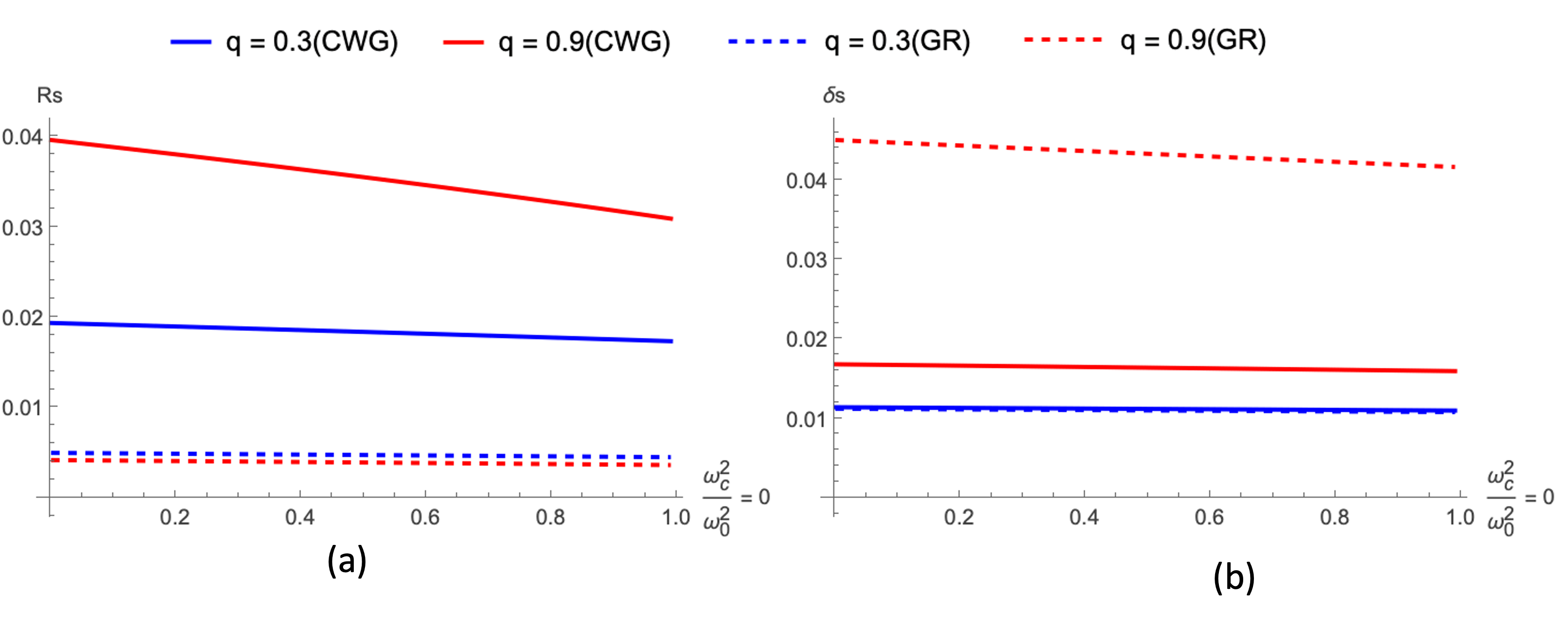}
    \caption{Finite-observer shadow observables as functions of plasma strength $\omega_c^2/\omega_0^2$. Solid curves denote CWG and dashed curves denote GR. Blue curves use $q=0.3$ and red curves use $q=0.9$. Panel (a) shows the effective radius $R_s$ and panel (b) shows the distortion $\delta_s$. We set $a=0.3$, $r_o=1000$, $\theta_o=\pi/2$, and $\Lambda=10^{-26}$.}
    \label{fig:Observablew}
\end{figure*}
The observable trends in Figs.~\ref{fig:Observableq} and \ref{fig:Observablew} quantify the visual behavior of Fig.~\ref{fig:Shadow-compare}. Figure~\ref{fig:Observableq} shows that increasing $q$ increases $R_s$ in CWG but decreases $R_s$ in GR. The difference becomes large at high charge, where the CWG charge term strongly modifies $\Delta_r$ and hence the finite-observer projection. Plasma lowers $R_s$ in both theories, but it does not change the sign of the charge dependence. The distortion $\delta_s$ increases with $q$ in both theories, although the GR distortion grows more rapidly near the upper end of the plotted charge range.

Figure~\ref{fig:Observablew} isolates the plasma dependence. For fixed charge, increasing $\omega_c^2/\omega_0^2$ decreases the effective radius in both CWG and GR. The decrease is mild at low charge and more visible for high-charge CWG, where the baseline shadow is larger. The distortion also decreases slightly as the plasma strength increases. At low charge the two theories predict similar distortions, but at high charge the GR distortion remains much larger than the CWG distortion. Taken together, these observables show that plasma modifies the shadow in both theories, while the charge dependence provides the stronger discriminator between the CWG and GR geometries.
\section{Energy Emission}
We next estimate the high-energy emission rate associated with the rotating charged CWG black hole immersed in plasma. The calculation is based on the usual geometric-optics approximation, in which the limiting absorption cross section approaches the capture cross section associated with the photon region and is represented, in shadow applications, by the apparent shadow area \cite{DecaniniFolacciRaffaelli2011,WeiLiu2013EMDAShadow}. In geometric units, the spectral emission rate is
\begin{equation}
    \frac{d^2Z(\omega)}{d\omega dt}
    =\frac{2\pi^2\sigma_{\rm lim}}{\exp(\omega/T_H)-1}\omega^3 ,
\end{equation}
where $\omega$ is the emitted radiation frequency, $T_H$ is the Hawking temperature, and $\sigma_{\rm lim}$ is the high-energy absorption cross section. We use dimensionless variables,
\begin{equation}
    \bar{\omega}=M\omega,\qquad
    \bar{T}_H=MT_H,\qquad
    \bar{\sigma}_{\rm lim}=\frac{\sigma_{\rm lim}}{M^2},
\end{equation}
so that, apart from the overall mass scaling, the plotted spectral shape is determined by
\begin{equation}
    \mathcal{F}(\bar{\omega})\equiv
    M\frac{d^2Z}{d\omega dt}
    =
    \frac{2\pi^2\bar{\sigma}_{\rm lim}}
    {\exp(\bar{\omega}/\bar{T}_H)-1}\bar{\omega}^3 .
\end{equation}
The radius entering the cross section is taken from the finite-observer shadow constructed in Sec.~V, rather than from Bardeen coordinates at infinity. If $A_{XY}$ is the area enclosed by the dimensionless screen curve $(X(r_p),Y(r_p))$, then
\begin{equation}
    R_{XY}=\sqrt{\frac{A_{XY}}{\pi}},\qquad
    A_{XY}=\frac{1}{2}\left|\oint_{\partial S}\left(X\,dY-Y\,dX\right)\right|.
    \label{eq:areaXY}
\end{equation}
For a distant finite observer whose radial position is measured in units of $M$, the dimensionless areal shadow radius is
\begin{equation}
    \bar{r}_{\rm sh}\equiv\frac{r_{\rm sh}}{M}
    \simeq r_o R_{XY}.
    \label{eq:dimensionless-rsh}
\end{equation}
This conversion is the finite-observer analogue of converting an angular shadow size into a physical radius. In the present dimensionless emission plots we therefore use
\begin{equation}
    \bar{\sigma}_{\rm lim}\simeq \pi \bar{r}_{\rm sh}^2 .
\end{equation}
The plasma is treated as a test medium, so it modifies the spectrum through the shadow cross section but does not change the horizon temperature.

\begin{figure*}
	\centering
    \includegraphics[width=\textwidth]{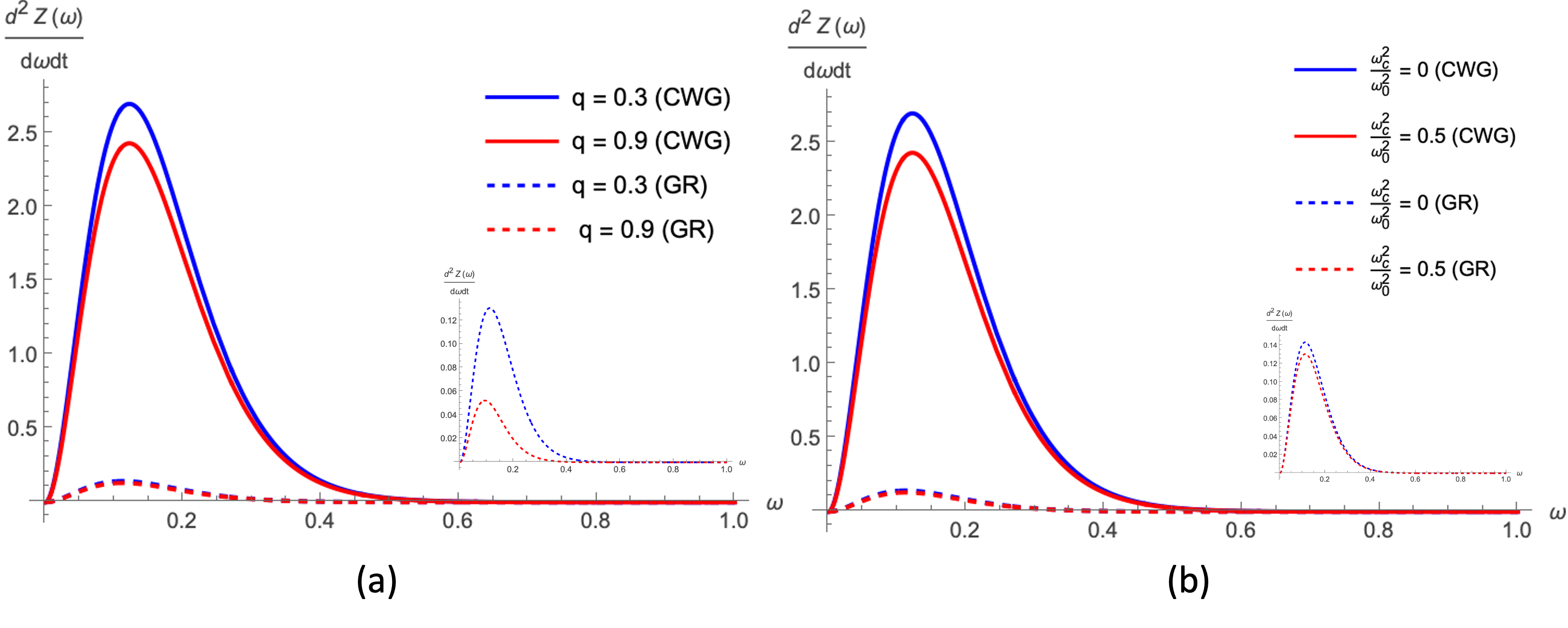}
    \caption{Dimensionless high energy emission rate for CWG (solid curves) and GR (dashed curves). Panel (a) varies charge, with $q=0.3$ in blue and $q=0.9$ in red, at fixed $a=0.1$, $\Lambda=10^{-26}$, and $\omega_c^2/\omega_0^2=0.5$, and $r_o = 1000$. Panel (b) varies plasma strength, with $\omega_c^2/\omega_0^2=0$ in blue and $\omega_c^2/\omega_0^2=0.5$ in red, at fixed $a=0.1$, $q=0.3$, and $\Lambda=10^{-26}$, $r_o = 1000$. The insets provide a magnified view of the dashed curves.}
    \label{fig:EM}
\end{figure*}

The Hawking temperature is computed from
\begin{equation}
    T_H=\frac{\kappa}{2\pi}.
\end{equation}
For the Kerr-like form of the metric, the surface gravity associated with the outer horizon generator is written as
\begin{equation}
    \kappa=\frac{\Delta_r'(r_+)}{2(r_+^2+a^2)} ,
    \label{eq:kappa-general}
\end{equation}
where $r_+$ is the outer black hole horizon determined by $\Delta_r(r_+)=0$. This form reduces to the usual Kerr expression when the radial function has the Kerr limit. 
Figure~\ref{fig:EM} shows how the resulting emission curves respond to charge and plasma. Panel (a) demonstrates that charge separates the two theories more strongly than plasma does. As the charge increases, the energy emission rate decreases in both theories, but the CWG curves remain higher than the corresponding GR curves. The separation grows with charge because the CWG radial function contains the charge contribution proportional to $q^2r^3/(6M)$, making the effect more pronounced at the finite observer radius used in the calculation. Panel (b) isolates the plasma dependence. Increasing $\omega_c^2/\omega_0^2$ lowers the emission peak in both theories, with a stronger suppression in CWG. Overall, charge and plasma both suppress the emission, but charge provides the sharper discriminator between CWG and GR.
\section{Joint Charge-Plasma Constraints from EHT-Inferred Shadow Sizes}
Horizon-scale images provide one of the few direct probes of strong-field gravity outside the weak-field and post-Newtonian regimes. In GR, the characteristic shadow radius is close to $3\sqrt{3}M$ for a nonrotating black hole, with corrections from spin, inclination, charge, plasma, and the observer position. CWG can alter this scale through both the photon region and the finite-observer projection. We therefore compare the predicted dimensionless shadow radius with the EHT-inferred intervals for Sgr~A$^*$ and M87$^*$, first at fixed charge and then across the joint charge-plasma parameter space \cite{EventHorizonTelescope:2019dse,EventHorizonTelescope:2022wkp,EHT2022VI,Vagnozzi:2023lwo}. The EHT image contains a bright emission ring, whereas our calculation gives the critical curve bounding the black hole shadow. Relating the two requires calibration with image and emission models \cite{EventHorizonTelescope:2022wkp,EHT2022VI}. Therefore, the constraints below should be interpreted as conditional geometric compatibility bounds for the adopted metric in both theories for a chosen plasma profile, spin, inclination, cosmological constant, and observer radius. 

For each model we compute the closed critical curve in the dimensionless finite-observer screen coordinates $(X,Y)$. Its area $A_{XY}$ gives the areal shadow radius
\begin{equation}
    \frac{r_{\rm sh}}{M}=r_o\sqrt{\frac{A_{XY}}{\pi}},
    \label{eq:model-rsh-over-m}
\end{equation}
where the polygonal area is evaluated from an ordered sampling of the complete shadow boundary, as defined in \ref{eq:areaXY}. The observational quantity is the fractional deviation from the Schwarzschild radius,
\begin{equation}\label{eq:rsh}
    \delta=\frac{r_{\rm sh}}{3\sqrt{3}M}-1,
    \qquad
    \frac{r_{\rm sh}}{M}=3\sqrt{3}(1+\delta).
\end{equation}
Using the averaged Keck/VLTI estimate $\delta=-0.060\pm0.065$ for Sgr~A$^*$ \cite{EHT2022VI,Vagnozzi:2023lwo}, and $\delta=-0.01\pm0.17$ at $68\%$ confidence for M87$^*$ \cite{Kocherlakota:2021dcv}, we adopt
\begin{align}
\text{Sgr A}^*:&&4.55&\leq r_{\rm sh}/M\leq5.22 &&(68\%),\nonumber\\
&&4.21&\leq r_{\rm sh}/M\leq5.56 &&(\text{approx. }95\%),\label{eq:eht-bands-sgr}\\
\text{M87}^*:&&4.26&\leq r_{\rm sh}/M\leq6.03 &&(68\%),\nonumber\\
&&3.38&\leq r_{\rm sh}/M\leq6.91 &&(\text{approx. }95\%).\label{eq:eht-bands-m87}
\end{align}
The quoted $95\%$ intervals are Gaussian extensions of the reported $68\%$ uncertainties and are used only as broader reference bands.
\begin{table}
\centering
\footnotesize
\setlength{\tabcolsep}{3.5pt}
\begin{tabular}{@{}lcc@{}}
\toprule
Theory & $\Delta_{r,{\rm charge}}$ & $\Delta_{r,{\rm charge}}/r^2$\\
\midrule
CWG & $q^2r^3/6$ & $q^2r/6$\\
GR  & $q^2$ & $q^2/r^2$\\
\bottomrule
\end{tabular}
\caption{Radial dependence of the charge terms for both theories. The cubic CWG term produces the finite-observer scaling $q\propto r_o^{-1/2}$, whereas the GR charge term has no radial dependence in $\Delta_r$.}
\label{tab:charge-scale}
\end{table}

\begin{figure*}[htbp]
  \begin{center}
    \includegraphics[width=\textwidth]{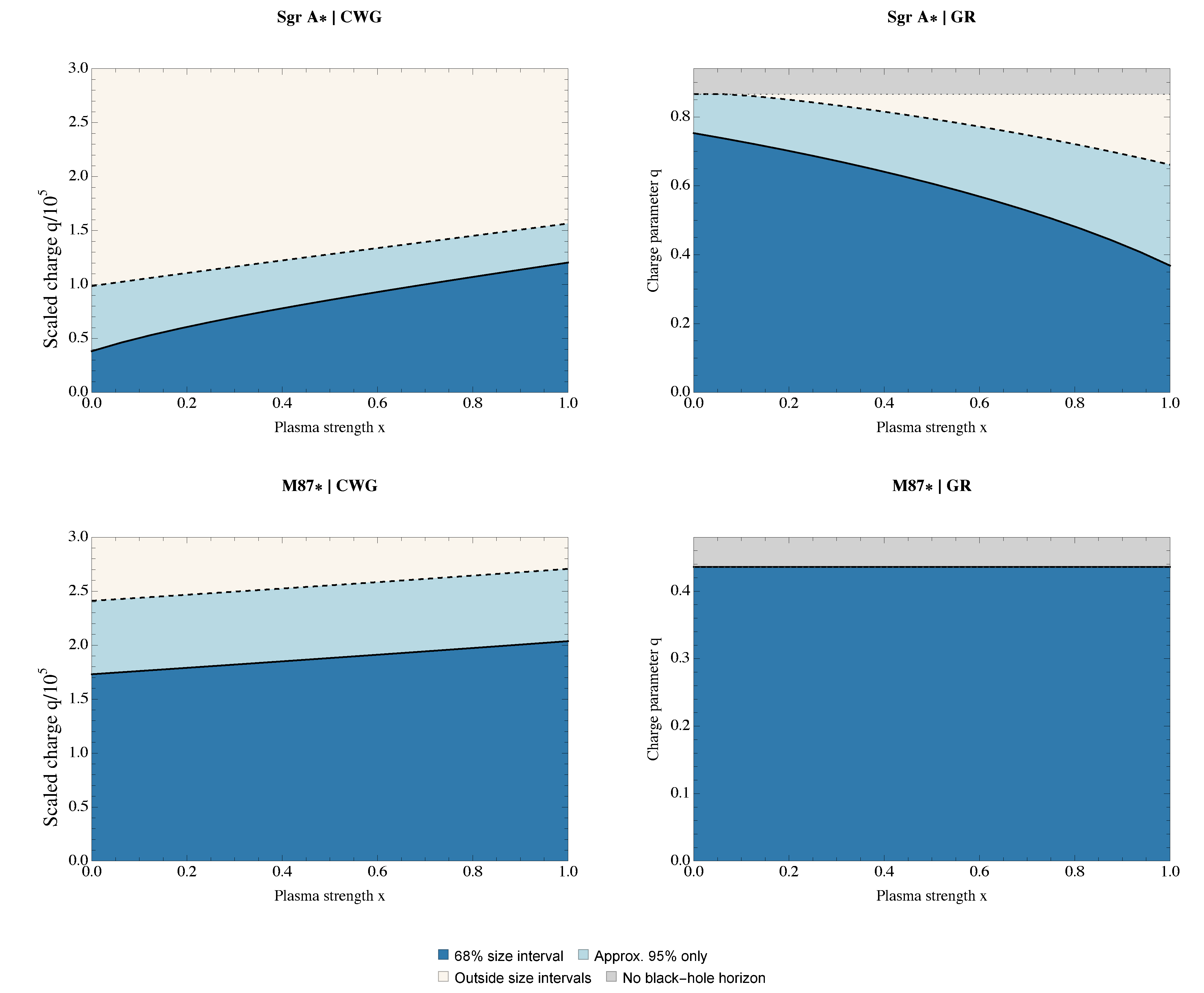}
     \caption{\label{fig:Constraint} Joint charge-plasma compatibility with the EHT-inferred shadow-size intervals for Sgr~A$^*$ (top row) and M87$^*$ (bottom row). The left and right columns show CWG and GR, respectively. Dark blue denotes compatibility with the $68\%$ interval, light blue denotes points compatible only with the approximate $95\%$ interval, beige denotes points outside both size intervals, and gray marks the absence of the required black-hole horizon. The solid and dashed curves are the $68\%$ and approximate $95\%$ charge boundaries; dotted lines mark the GR horizon limit. Each column scans the charge parameter of its own metric over a theory-specific range; the CWG vertical axes display $10^5/q$, whereas the GR axes display $q$. These axes do not represent a common physical-charge normalization. We use $r_o=10^{10}$ and $\Lambda=10^{-26}$, with $(a,\theta_o)=(0.5,\pi/2)$ for Sgr~A$^*$ and $(0.9,17^\circ)$ for M87$^*$.}
 \end{center}
\end{figure*}

We write the dimensionless plasma strength as
\begin{equation}
    x=\frac{\omega_c^2}{\omega_0^2}, \qquad 0\leq x\leq1,
\end{equation}
and denote either theory by $T=\mathrm{CWG}$ or $T=\mathrm{GR}$. At fixed $x$, a charge is compatible with an observational band $[L_{n\sigma},U_{n\sigma}]$  ($n\sigma$ means either $68\%$ or $95\%$ confidence interval) when
\begin{equation}
 L_{n\sigma}\leq\frac{r_{\rm sh}^{T}(q,x)}{M}\leq U_{n\sigma}.
 \label{eq:joint-compatibility}
\end{equation}
This condition gives the charge bound directly. Since the numerical scan verifies that $r_{\rm sh}^{\rm CWG}$ increases monotonically with $q$, whereas $r_{\rm sh}^{\rm GR}$ decreases monotonically with $q$, as shown in Fig.~\ref{fig:Constraint}, we obtain the CWG size bound from the upper observational edge, and the GR size bound from the lower edge. The maximum allowed charge $q_{\rm max}$ is the largest charge for which the required horizon structure remains present. We take $r_o=10^{10}$ and $\Lambda=10^{-26}$. For Sgr~A$^*$ we use  $a=0.5$ and $\theta_o=\pi/2$; the equatorial inclination is a calculation benchmark rather than the EHT-preferred orientation \cite{Kato2009SgrASpin,EventHorizonTelescope:2022wkp}. For M87$^*$ we use $a=0.9$ and $\theta_o=17^\circ$ \cite{Tamburini2020,Nemmen2021}. Current spin estimates remain model dependent, so the resulting bounds are conditional on these choices \cite{Johnson2023ngEHT}. 
We treat the parameter scans in the following ways for the two theories. For the charge bounds, we fix $x$ and vary $q$ independently in each theory. Their different numerical ranges follow from the radial charge terms summarized in Table~\ref{tab:charge-scale}. With $M=1$, the CWG charge term in $\Delta_r$ is $q^2r^3/6$. Since the metric contains $\Delta_r/\rho^2$ and $\rho^2\simeq r^2$ at the distant observer, the corresponding large-radius contribution scales as $q^2r_o/6$. For $r_o=10^{10}$, $q=10^{-5}$ gives $q^2r_o/6=0.167$, while a unit contribution would occur at $q=\sqrt{6/r_o}=2.45\times10^{-5}$. This sets the natural $10^{-5}$ scale of the CWG charge scan. In GR, the charge term in $\Delta_r$ is $q^2$, and its metric contribution is $q^2/r^2$. Near the strong-field region, where $r=O(1)$, choosing $q=10^{-5}$ therefore gives only an $O(10^{-10})$ correction and is numerically indistinguishable from the neutral Kerr-de Sitter limit. We consequently scan $q$ over its order-unity range, up to the horizon-extremality limit, to resolve the Kerr-Newman-de Sitter charge effect. For the plasma bounds, we also show the shared numerical slice $q=10^{-5}$ in both metrics; its GR result serves as the neutral baseline.
In contrast to the asymptotically flat Kerr-Newman case, where the charge contribution at the observer decreases as $q_{\rm GR}^2/r_o^2$, the charged CWG geometry has a stronger finite-observer dependence. For $r_o\gg M$, with $\rho_o^2\simeq r_o^2$, the CWG charge term contributes $q_{\rm CWG}^2r_o/(6M)$ to $\Delta_r/\rho_o^2$. When this term controls the observer-radius dependence, holding the shadow interval fixed requires $q_{\rm CWG}^2r_o$ to remain approximately constant, giving
\begin{equation}
q_{\rm max}^{\rm CWG}\propto r_o^{-1/2}.
\end{equation}
\onecolumngrid
\par\medskip
\begin{table}
\centering
\setlength{\tabcolsep}{5.2pt}
\begin{tabular}{@{}llccccc@{}}
\toprule
Source & Theory & EHT interval & Allowed $x$ at shared $q=10^{-5}$
& $q_{\rm max}(0)$ & $q_{\rm max}(0.5)$ & $q_{\rm max}(1)$\\
\midrule
Sgr~A$^*$ & CWG & $68\%$ & $[0.698,1]$ & $3.83\times10^{-6}$ & $8.57\times10^{-6}$ & $1.20\times10^{-5}$\\
             & CWG & approx. $95\%$ & $[0.022,1]$ & $9.87\times10^{-6}$ & $1.28\times10^{-5}$ & $1.57\times10^{-5}$\\
             & GR  & $68\%$ & $[0,1]$ & $0.753$ & $0.607$ & $0.368$\\
             & GR  & approx. $95\%$ & $[0,1]$ & $0.866^{\dagger}$ & $0.795$ & $0.661$\\
\midrule
M87$^*$      & CWG & $68\%$ & $[0,1]$ & $1.73\times10^{-5}$ & $1.88\times10^{-5}$ & $2.04\times10^{-5}$\\
             & CWG & approx. $95\%$ & $[0,1]$ & $2.41\times10^{-5}$ & $2.55\times10^{-5}$ & $2.71\times10^{-5}$\\
             & GR  & $68\%$ & $[0,1]$ & $0.436^{\dagger}$ & $0.436^{\dagger}$ & $0.436^{\dagger}$\\
             & GR  & approx. $95\%$ & $[0,1]$ & $0.436^{\dagger}$ & $0.436^{\dagger}$ & $0.436^{\dagger}$\\
\bottomrule
\end{tabular}
\caption{Combined plasma and charge compatibility bounds from the EHT-inferred shadow sizes. The fourth column varies $x=\omega_c^2/\omega_0^2$ while holding the same numerical benchmark $q=10^{-5}$ in both metrics. The three rightmost columns instead vary $q$ independently in each theory at fixed $x$ and give the corresponding upper bounds; all charge intervals begin at $q=0$. These three columns correspond to different plasma strength ($\frac{\omega_c^2}{\omega_0^2} = 0, 0.5, 1$). A dagger marks a horizon-extremality bound, and all unmarked entries are limited by the shadow-size interval. The M87$^*$ $95\%$ interval is an approximate Gaussian extension. Because the CWG and GR charge parameters have different normalizations and radial couplings, their numerical bounds are theory-specific conditional geometric limits, not posterior credible intervals on a common physical electric charge.}
\label{tab:constraint}
\end{table}
\twocolumngrid
Figure~\ref{fig:Constraint} displays the joint compatibility regions. Plasma makes the shadow smaller in both theories, but charge does not: it enlarges the CWG shadow and shrinks the GR shadow, which is consistent with the results shown in Fig.~\ref{fig:Observablew}[a]. In CWG, plasma therefore offsets the effect of charge, so larger plasma strength $x$ allows larger $q$ and the boundary slopes upward. In GR, plasma and charge work in the same direction, so larger $x$ lowers $q_{\rm max}^{\rm GR}$ when the shadow-size limit is active, giving a downward-sloping boundary. The M87$^*$ GR panels are an exception because the observational interval is broad enough that the horizon condition, rather than the shadow size, sets the upper bound. The opposite slopes, not the very different numerical scales of $q$, are the main test provided by the figure.

Table~\ref{tab:constraint} combines the two one-parameter readings of the joint map. At the fixed charge $q=10^{-5}$, the Sgr~A$^*$ CWG curve requires $x\geq0.698$ for the $68\%$ interval and $x\geq0.022$ for the approximate $95\%$ interval; all other listed fixed-charge plasma ranges span $0\leq x\leq1$. The same table gives $q_{\rm max}$ at three representative plasma strengths. For example, the Sgr~A$^*$ $68\%$ CWG bound increases from $3.83\times10^{-6}$ in vacuum to $1.20\times10^{-5}$ at $x=1$, whereas the GR bound decreases from $0.753$ to $0.368$. 
\section{Summary \& Conclusions}\label{sec:5}
We studied photon propagation, finite-distance lensing, shadow formation, and geometric-optics emission for a charged rotating de Sitter black hole in fourth-order conformal Weyl gravity surrounded by a cold, non-magnetized plasma. Its charge contribution enters the radial function as $q^2r^3/(6M)$ and therefore has a different radial dependence from the charge term in the standard Kerr-Newman-de Sitter spacetime.

The horizon analysis identified the parameter domain in which the event horizon remains distinct from the Cauchy and cosmological horizons. That analysis also explains why charge affects the optical observables differently in the two theories. After introducing a separable plasma profile in the Hamilton-Jacobi equation, we obtained radial and angular potentials in which the same functions $f_r(r)$ and $f_\theta(\theta)$ are weighted by different metric functions. Plasma is therefore not an additive correction independent of the geometry.

The finite-distance deflection angle $\alpha$ gives an independent lensing check of this result. For equatorial rays with $r_S=20$, $r_O=1000$, $a=0.8$, and $\Lambda=10^{-26}$, increasing the impact parameter lowers the signed angle along each plotted branch. Increasing either $q$ or $\omega_c^2/\omega_0^2$ also shifts $\alpha$ downward. The GR curves remain positive and depend only weakly on charge over the displayed range, whereas the CWG curves decrease rapidly with $q$ and cross zero at sufficiently high charge. Plasma strengthens the downward shift, especially in the high-charge CWG case. These lensing results agree with the earlier photon-region analysis: the plasma response occurs in both theories, but the charge response separates CWG from GR more clearly.

Because both spacetimes are not asymptotically flat, we constructed the shadow on the local sky of a finite observer and mapped the observation angles to the screen coordinates $(X,Y)$. Increasing plasma strength decreases the effective shadow radius in both theories. Charge produces the more distinctive trend for the chosen parameters: the CWG shadow grows with $q$, whereas the GR shadow shrinks. The radius $R_s$ and distortion $\delta_s$ quantify this difference. The geometric-optics emission estimate follows the same pattern. Plasma lowers the emission peak in both theories, while charge produces a larger CWG-GR separation; at high charge, the plotted CWG spectrum remains above the corresponding GR spectrum.

For the EHT comparison, we selected the model parameters compatible with the published shadow-size intervals. The joint $(q,\omega_c^2/\omega_0^2)$ maps shows an opposite degeneracy between charge and plasma. Plasma reduces the shadow radius in both theories; it therefore offsets the CWG shadow enlargement caused by charge but reinforces the GR shadow reduction. As a result, stronger plasma raises the CWG charge bound and lowers the GR charge bound whenever shadow size is limited by the horizon extremality. For the Sgr~A$^*$ $68\%$ interval, maximum allowed charge in CWG increases from $3.83\times10^{-6}$ in vacuum to $1.20\times10^{-5}$ at $\omega_c^2/\omega_0^2=1$, while maximum allowed charge in GR decreases from $0.753$ to $0.368$. At the fixed charge $q=10^{-5}$, CWG requires $\omega_c^2/\omega_0^2\geq0.698$ for the Sgr~A$^*$ $68\%$ interval and $\omega_c^2/\omega_0^2\geq0.022$ for the approximate $95\%$ interval. The broader M87$^*$ interval leaves GR limited by horizon extremality over the full plasma range, while CWG remains size limited.

This work extends earlier plasma-free shadow and EHT studies of the charged rotating CWG geometry by including a separable plasma distribution and joint charge-plasma bounds for both EHT sources. Our key result is the opposite trend of the charge boundary in CWG and GR where shadow size sets the limit. We show how the distinctive CWG radial charge term impacts the measurable shadow property and how the inferred charge range changes with the assumed plasma strength. The present calculations use only $r_o=10^{10}$ as the radial location of the observer. A possible extension to this work would be to survey the observer radii needed to test this scaling and quantify its effect on the inferred CWG charge bounds due to the finite-observer dependence from the cubic radial charge term. \\

\begin{acknowledgments}
We acknowledge support from the Grinnell College CSFS grant for this work.
\end{acknowledgments}

\bibliographystyle{apsrev4-2} 
\bibliography{references.bib}

\newpage
\newpage

\end{document}